\documentclass[letterpaper,journal]{IEEEtran}
\usepackage[utf8]{inputenc}
\usepackage{amsmath,amsfonts}
\usepackage{algorithmic}
\usepackage{algorithm}
\usepackage{array}
\usepackage[caption=false,font=normalsize,labelfont=sf,textfont=sf]{subfig}
\usepackage{textcomp}
\usepackage{stfloats}
 \usepackage{url}
\usepackage{verbatim}
\usepackage{graphicx}
\usepackage{cite}
\usepackage{verbatim}
\usepackage{multirow}
\usepackage{orcidlink}
\usepackage{verbatim}

\begin{document}

\title{TREA: Low-precision \textbf{T}ime-Multiplexed, \textbf{R}esource-Efficient \textbf{E}dge \textbf{A}ccelerator for Object Detection and Classification}

\author{Vijay Pratap Sharma, Mukul Lokhande~\textsuperscript{\orcidlink{0009-0001-8903-5159}},~\IEEEmembership{Member, IEEE},\\ Ratko Pilipović \textsuperscript{\orcidlink{0000-0002-4346-5487}}, Omkar Kokane~\textsuperscript{\orcidlink{0009-0000-6288-7231}},
Santosh Kumar Vishvakarma~\textsuperscript{\orcidlink{0000-0003-4223-0077}},~\IEEEmembership{Senior Member, IEEE} 
        % <-this % stops a space

\thanks{Manuscript received ; revised .}
\thanks{Vijay Pratap Sharma, Omkar Kokane, Mukul Lokhande and Santosh Kumar Vishvakarma are with the NSDCS Research Group, Department of Electrical Engineering, Indian Institute of Technology Indore, Indore, India.\\
Ratko Pilipovic is with the Faculty of Computer and Information Science, University of Ljubljana, Ljubljana, Slovenia.\\
This work was supported by the Special Manpower Development Program
for Chip to Start-Up (SMDP-C2S), Ministry of Electronics and Information
Technology (MeitY), Government of India, under Grant EE-9/2/21 - R\&D-E.\\
\textbf{Corr. author}: Dr. Santosh K. Vishvakarma, \textbf{E-mail:} skvishvakarma@iiti.ac.in}}

% The paper headers
\markboth{IEEE Transactions on Very Large Scale Integration (VLSI) Systems, ~Vol.~XX, No.~X, XXX~202X}%
{Sharma \MakeLowercase{\textit{et al.}}: Time-multiplexed, Resource-efficient Edge-AI accelerator for deep neural networks}

\maketitle

\begin{abstract}
This work presents TREA, a low-precision time-multiplexed and resource-efficient edge-AI accelerator for object detection and classification, targeting stringent area-power-latency constraints of edge vision platforms. The proposed architecture integrates a dual-precision (4/8-bit) SIMD multiply-accumulate (DQ-MAC) unit based on most-significant-digit-first (MSDF) shift-and-add computation with run-time bit truncation, eliminating conventional multiplier overhead and reducing accumulator bit-width. The DQ-MAC supports 4x FxP4 or 1x FxP8 operations per cycle, achieving up to 4x throughput improvement without hardware duplication. A structured hardware-aware reductive pruning (SHARP) strategy is co-designed with the SIMD datapath, enabling near 50\% structured sparsity while maintaining full MAC utilization. This allows a 3x3 convolution kernel to be computed in 1 cycle in FxP4 mode compared to 9 cycles in FxP8, and a 5x5 kernel in 3 cycles versus 25 cycles, yielding up to 9x latency reduction at the kernel level. The accelerator further incorporates a reconfigurable CORDIC-based nonlinear activation function (RQ-NAF) core with a 9-stage pipeline, supporting Sigmoid, Tanh, and ReLU at one output per cycle after pipeline fill, while enabling (N-1) hardware reuse through time-multiplexing. The complete TREA architecture employs a 1D array of 100 SIMD DQ-MAC units with layer-wise hardware reuse, significantly reducing area and control complexity. Experimental results demonstrate substantial improvements in latency, hardware utilization, and energy efficiency compared to conventional fixed-precision and non-reconfigurable accelerators, validating TREA as an effective solution for real-time edge vision workloads.
\end{abstract}


\section{Introduction}

\IEEEPARstart{F}{rom} eye tracking and object detection to real-time scene recognition, machine perception algorithms form the computational core of Edge Vision and Visual AI~(EV2AI) systems deployed on resource-constrained embedded platforms. EV2AI models are simultaneously compute- and memory-intensive: the ViT-G object detection model~\cite{zhai2022scaling} alone requires 2.86~GMACs and 184~billion parameters per inference. Memory access dominates energy consumption, ranging from approximately 5~pJ (on-chip SRAM) to 640~pJ (off-chip DRAM) in 45~nm CMOS~\cite{LWC_QP}, such that a 1-billion-parameter model at 20~Hz demands approximately 12.8~W for DRAM accesses alone, well beyond the sub-watt budgets of edge platforms. Their computational complexity imposes a fundamental trilemma among accuracy, latency, and energy consumption~\cite{XRTrilemma'24}, particularly for battery-powered devices.

 %Aggressive parameter optimization, quantization, and connection pruning are therefore essential to viable deployment: eliminating redundant kernel elements reduces the active parameter count, while lowering numerical precision jointly decreases parameter storage requirements, off-chip data movement, and the energy associated with inactive neuron computation~\cite{PvQ_QC}.
%Apple's Vision Pro, for instance, achieves only two hours of battery life, compared to Meta's Quest Pro, which requires approximately $1.7\times$ higher battery capacity and $3.4\times$ greater battery weight for comparable operation. This energy inefficiency largely stems from insufficient exploitation of the inherent error resilience of machine perception algorithms~\cite{XR_comp}, particularly through hardware-aware execution strategies that account for approximately 60-80\% of energy-efficient machine perception acceleration~\cite{Aspen_CICC'25}.

The optimal design of EV2AI accelerators demands high throughput, compactness, and energy efficiency within strict area and power budgets. From a hardware perspective, EV2AI accelerators require extensive spatial parallelism, structured dataflow, and tailored memory hierarchies to efficiently support matrix- and vector-centric operations~\cite{Flex-PE}. The foundational compute blocks comprise multiply-accumulate (MAC) units for matrix operations and nonlinear activation functions (NAFs) such as ReLU, Sigmoid, and Tanh. To further improve the efficiency of EV2AI accelerators, model-level optimisations must be considered alongside hardware design. Precision scaling reduces memory footprint and off-chip data movement by representing weights and activations at lower bitwidths \cite{FALCON, VLSID'26}, while pruning \cite{Zhu-SparseCNN-TVLSI20} eliminates redundant computations, enabling mixed precision across network layers.

\begin{comment}
\begin{table}[!t]
\centering
\caption{State-of-the-art AI Models with Parameter Count, MAC Complexity, Sparsity Trends projecting the opportunity for hardware optimisation - Could be ommited }
\label{tab:model_parameters}
\renewcommand{\arraystretch}{1.15}
\resizebox{\columnwidth}{!}{%
\begin{tabular}{l|c|c|c|c|c}
\hline
\textbf{Model} &
\textbf{Dataset} &
\textbf{Parameters} &
\begin{tabular}[c]{@{}c@{}}\textbf{MACs}\\ \textbf{(Billions)}\end{tabular} &
\textbf{Sparsity (\%)} &
\begin{tabular}[c]{@{}c@{}}\textbf{Physical}\\ \textbf{Overhead}\end{tabular} \\ 
\hline
VGG-16 & ImageNet & 138M & 15.5 & 80 & Medium \\ \hline
ResNet-50 & ImageNet & 25.6M & 3.8 & 65 & High \\ \hline
MobileNet-v1 & ImageNet & 4.2M & 0.57 & 35 & Low \\ \hline
MobileBERT & SST-2 & 25.3M & 4.3 & 65 & Low \\ \hline
Inception-v3 & ImageNet & 23.9M & 5.7 & 60 & High \\ \hline
BERT-Base & SST-2 & 110M & 21.2 & 55 & Medium \\ \hline
GPT-2 (Large) & SST-2 & 117M & 42 & 35 & Medium \\ \hline
GPT-3 & Internet-scale & 175B & 6700 & 75 & Very High \\ \hline
GPT-4* & Internet-scale & 1700B\textsuperscript{$\dagger$} & $>100000$ & 50 & Extreme \\ \hline
\end{tabular}}
\end{table}
\end{comment}

Majority of works largely target isolated levels of the design stack, addressing quantization, pruning, MAC unit design, or architectural reuse independently, without a unified co-design framework spanning the full algorithm-to-hardware stack. In this work, all of the levels in the design stack are addressed in such manner that mutually reinforce one another to yield substantial improvements in hardware uttization and energy-efficiency. The key contributions of this work are summarized as follows: 

\begin{itemize}
    \item \textbf{DQ-MAC:} {We propose a multiplier-free dual-precision (4/8-bit) SIMD MAC unit based on MSDF shift-and-add computation. Unlike conventional multiplier-based or post-quantization MAC designs, the proposed DQ-MAC performs pre-accumulation truncation via MSD-guided power-of-two decomposition, thereby reducing accumulator bit-width, eliminating post-multiplier quantization hardware, and enabling up to 4$\times$ throughput without hardware duplication.}

    \item \textbf{SHARP:} {We introduce SHARP, a structured hardware-aware pruning strategy that co-designs sparsity with the SIMD datapath. By enforcing kernel-wise weight retention aligned to the four-lane SIMD width (4:9 and 12:25 patterns), SHARP guarantees full lane utilisation, eliminates partially occupied execution cycles, and transforms pruning into a hardware-efficient scheduling mechanism.}

    \item \textbf{RQ-NAF:} {We design a reconfigurable CORDIC-based nonlinear activation core supporting Sigmoid, Tanh, and ReLU within a unified pipelined datapath. A single shared instance is time-multiplexed across layers, significantly reducing activation hardware overhead while sustaining one output per cycle after pipeline fill.}

    \item \textbf{TREA Architecture:} {We develop a time-multiplexed 1D accelerator architecture that integrates the DQ-MAC array, SHARP pruning, and RQ-NAF core under a unified control engine. The design enables layer-wise hardware reuse, reduces control complexity, and supports scalable deployment across varying DNN topologies. {An algorithm–hardware co-design framework that jointly optimises SIMD-aligned structured pruning, quantisation, MAC execution, activation reuse, and time-multiplexed scheduling for efficient edge object detection and classification.}}

    \item \textbf{End-to-End Validation:} {We demonstrate the effectiveness of TREA through comprehensive evaluation, including FPGA implementation and 28-nm ASIC synthesis and place-and-route, along with object detection and classification benchmarks, showing improvements in latency, energy efficiency, and resource utilisation over state-of-the-art designs.}
\end{itemize}

\section{Background \& Motivation}

\subsection{Efficient inference of EV2AI models}

For EV2AI inference, quantization reduces memory footprint and computational complexity by representing weights and activations in lower-precision fixed-point or floating-point formats, enabling low-bitwidth arithmetic that significantly reduces energy consumption and off-chip memory traffic. Determining the optimal bitwidth per layer remains non-trivial~\cite{HAQ}, and various schemes have been proposed to balance accuracy and hardware efficiency, ranging from group quantization with independent per-group scaling factors~\cite{Qserve} to power-of-two representations~\cite{poweroftwo} that reduce computational cost at the expense of accuracy. Accuracy degradation introduced by quantization can be mitigated through quantization-aware training~\cite{Smoothquant} and fine-tuning~\cite{SVDQuant}.

Orthogonal to quantization, pruning reduces latency and energy by eliminating redundant weights and neurons, thereby sparsifying the model and reducing the total operation count~\cite{Lu-SparseCNN-TCASI22, POLARON}. However, unstructured sparsity often incurs significant indexing overhead and irregular memory access patterns, diminishing hardware efficiency~\cite{Tiled-ECG}. Structured pruning~\cite{Yin-StrucSparse-ESL24, 9034111} addresses this by enforcing regular sparsity patterns that align naturally with hardware dataflow and processing element utilization, offering a favorable tradeoff between compression ratio and execution regularity.

\subsection{MAC unit design}

In MAC unit design, two principal processing paradigms exist: pipelined~\cite{FMA_TCAS-I'25} and serial execution~\cite{QuantMAC}, each exposing distinct trade-offs between throughput, area, and power consumption. For a MAC operation over $K$ kernel elements, each represented as an $N$-bit integer, the accumulator must be at least $2N + \lceil \log_2 K \rceil$ bits wide to prevent overflow. This bit-width expansion significantly increases compute complexity, data movement, and NAF unit hardware requirements, constraining system-level performance~\cite{QuantMAC, ART-MAC, QForce-RL}. Existing approaches mitigate accumulator growth through post-accumulation quantization prior to NAF units, albeit at non-trivial cost in hardware complexity and latency.

A further distinction exists between bit-serial or bit-parallel computation at the digit-level. In least-significant-digit-first (LSDF) arithmetic, all digits must be processed before a valid output is produced, requiring full-precision computation regardless of the operand magnitude. In contrast, most-significant-digit-first (MSDF) arithmetic processes operands from the most significant digit, generating the first valid output after $N$ cycles for an $N$-bit operand and producing one additional output digit per cycle thereafter. The proposed DQ-MAC exploits MSDF processing to prioritise the most information-carrying digits early, selectively bypassing lower-significance bits to reduce carry-propagation delay and improve resource efficiency through digit-level parallelism.

Approximate computing reduces MAC unit area and power by targeting multiplier design. Logarithmic multipliers convert multiplication into addition~\cite{ILM-AA}, while non-logarithmic alternatives reduce partial products through approximate encoding~\cite{RAD1024} and employ approximate compressors~\cite{10892007} for more efficient accumulation. An alternative to multiplier-based designs is the CORDIC algorithm, which realizes arithmetic operations through iterative shift-and-add rotations, replacing power-hungry multipliers with area-efficient logic. Reconfigurable CORDIC architectures support both multiplication and division~\cite{7152967}, and have been extended to performance-centric MAC units~\cite{RECON}, making them compelling candidates for edge accelerators.

%SIMD-based processing elements further leverage low-precision arithmetic by processing multiple data elements within a single instruction cycle, improving throughput without increasing memory bandwidth demands~\cite{GR-ISQED24}, while dual-precision support enabling runtime selection between 4-bit and 8-bit fixed-point modes offers additional flexibility to balance accuracy and efficiency on a per-layer basis~\cite{Flex-PE}.

\subsection{Non-linear Activation Functions (NAF)}

The EV2AI models owe their success to the diversity of employed NAFs, that need to be addressed by efficient accelerator architecture. 
Hardware implementations of NAF can consume up to 25\% of the total accelerator area~\cite{TPU-V4}. Existing NAF unit landscape features a broad range of architectural approaches, aimed for various applications. The first group leverages Taylor-series~\cite{TCASII-23_ReAFM, TCAD19-AF}, which can support broad range of activation functions operating at reduced bitwidths (7-12 bits) with fixed precision. However, the main drawback as they remain limited to a single processing element and lack reconfigurability. 

The CORDIC algorithm represents a compelling alternative as it can directly approximate the exponential function, which underlies all commonly employed NAFs. CORDIC-based designs~\cite{TAI24-CORDIC-RNN, TC23-CORDIC-LSTM} support Sigmoid and Tanh at fixed 16/24-bit precision in pipelined architectures, optimised for area efficiency in RNN and LSTM workloads.  Multi-precision and multi-format designs~\cite{MRao-ISQED24} broaden numerical support to BF16, TF32, Posit, and standard floating-point, targeting edge inference with parallel and pipelined execution, but without runtime reconfigurability. Among the most recent works, Flex-PE~\cite{Flex-PE} supports Sigmoid, Tanh, SoftMax, and ReLU across 4/8/16/32-bit fixed precision, targeting a wide range of AI workloads with a runtime area-latency trade-off.

\subsection{Hardware reuse in accelerator design}

Hardware accelerators for EV2AI inference must balance throughput, memory efficiency, and power consumption under strict area budgets. Fully parallel architectures~\cite{Fully-Parallel} achieve high throughput but incur prohibitive area and power costs for edge deployment. Hardware reuse strategies address this by dynamically sharing computation units across DNN layers, extending to activation functions and fully connected layers~\cite{9352042}. Time-multiplexed architectures further systematise resource efficiency by executing all DNN stages on a single configurable hardware instance~\cite{GR-Neuro, HYDRA}, reducing static power, minimising redundant logic, simplifying inter-layer data dependencies, and enabling a favorable area-throughput tradeoff under tight silicon area constraints.

\subsection{Motivation}

Although several prior accelerators address low-precision arithmetic, CORDIC-based computation, or structured pruning independently, TREA jointly optimizes the arithmetic datapath, pruning pattern, activation reuse, and layer-wise execution schedule within a unified edge-AI accelerator. At the MAC level, Flex-PE~\cite{Flex-PE} and QuantMAC~\cite{QuantMAC} improve computational efficiency through complementary means: the former via multi-precision SIMD execution, the latter via quantization-enabled accumulation. In contrast, TREA employs a multiplier-free MSDF shift-and-add DQ-MAC with pre-accumulation truncation, embedding quantization directly within the arithmetic datapath rather than treating it as an external precision mode or a post-multiplier operation. Regarding arithmetic representation, TREA targets fixed-point 4/8-bit SIMD execution, avoiding the posit-decoding and regime-handling overhead incurred by PQRE~\cite{pqre}, making the DQ-MAC more suitable for compact object-detection workloads. At the pruning level, unlike prior structured pruning approaches~\cite{9034111,Zhu-SparseCNN-TVLSI20}, which predominantly remove filters, channels, or blocks without regard to datapath alignment, SHARP explicitly maps the pruning pattern to the SIMD lane organization, ensuring that retained weights match the four-lane execution width and preventing lane under-utilization. At the activation level, TREA employs CORDIC exclusively within a shared reconfigurable nonlinear activation unit (RQ-NAF), while a separate SIMD DQ-MAC array handles convolutional computation; this contrasts with RECON~\cite{RECON}, which relies on CORDIC as its primary neuron-level compute primitive. By integrating these elements within a single architecture, TREA provides layer-wise time-multiplexed execution and activation reuse rather than optimizing a single compute primitive in isolation.

Therefore, the novelty of TREA lies in the following integrated contributions:
\begin{itemize}
    \item a multiplier-free MSDF DQ-MAC with embedded pre-accumulation truncation;
    \item a four-lane 4/8-bit SIMD datapath that improves throughput without duplicating hardware;
    \item SHARP, a SIMD-aligned structured pruning method that guarantees full lane utilization;
    \item a shared CORDIC-based RQ-NAF unit for runtime-selectable activation functions;
    \item a time-multiplexed 1D accelerator architecture combining DQ-MAC, SHARP, and RQ-NAF for object detection and classification.
\end{itemize}

\section{Proposed contribution}

\subsection{MSD-guided Power-of-Two Quantisation via Run-Time Bit-Truncation (MSD-PoTQ8)}

%The proposed DQ-MAC unit combines two complementary mechanisms to address accumulator bit-width growth and computational overhead. Dynamic power-of-two (PoT) quantization at runtime shifts quantization to the pre-accumulation stage, directly reducing the required accumulator width and eliminating post-accumulation quantization hardware. Concurrently, MSDF processing prioritizes the most information-carrying digits early, selectively bypassing lower-significance bits to reduce carry-propagation delay and improve resource efficiency and energy savings through digit-level parallelism. The combination of these two principles gives rise to the proposed quantization scheme.

For a normalized weight $|W_{\text{in}}| < 1$, the proposed scheme approximates $W_{\text{in}}$ using a finite signed power-of-two (PoT) expansion. At the $i$-th iteration, the residual weight is denoted by $W_i$, with $W_0 = W_{\text{in}}$. The power-of-two quantization term and the residual update are given, respectively, by \begin{equation} q_i = \mathrm{sign}(W_i) \, 2^{-\lfloor \log_2 |W_i| \rfloor}, \qquad W_{i+1} = W_i - q_i. \label{eq:dqmac_residual} \end{equation}
After $T$ iterations, the weight approximation is given by
\begin{equation}
\widehat{W}_{T} = \sum_{i=0}^{T-1} q_i,
\end{equation}
where each term $q_i$ corresponds to a signed power-of-two component ordered from the most to the least significant digit, which motivates the name of the proposed scheme as MSD-guided power-of-two quantization.

Using an $N$-bit signed fixed-point representation with $F$ fractional bits, the product of $X_{\text{in}}$ and $W_{\text{in}}$ is approximated by accumulating $T$ signed power-of-two  terms. To preserve the output word length and prevent accumulator growth, each shifted term is truncated prior to accumulation, yielding
\begin{equation} \widehat{Y}_{T} = \sum_{i=0}^{T-1} s_i \cdot \mathrm{trunc}_{N}\!\left(X_{\text{in}} \, 2^{-m_i}\right). \label{eq:dqmac_output}
\end{equation}
Fig. 2 illustrates the proposed MSD-guided multiplication using signed fixed-point arithmetic.

The proposed multiplication induces error multiplication error consists of two components: the residual weight approximation error and the truncation error. Hence,
\begin{equation}
\left|
X_{\text{in}}W_{\text{in}} - \widehat{Y}_{T}
\right|
\leq
|X_{\text{in}}| \cdot |W_T|
+
T\cdot2^{-F}.
\label{eq:total_error}
\end{equation}
Equation \eqref{eq:total_error} shows that increasing the number of MSD iterations reduces the residual approximation error, while the truncation error grows linearly with the number of accumulated truncated terms. This motivates the selected pipeline depth, which provides a favorable tradeoff between accuracy, latency, and hardware cost. 

At every iteration, the selected term $q_i$ corresponds to the most significant power-of-two component of the residual, ensuring monotonic convergence, i.e., $|W_{i+1}| = |W_i - q_i| < |W_i|$. For an $N$-bit fixed-point representation with $F$ fractional bits, the smallest representable step is $2^{-F}$; consequently, the residual converges to zero after at most $F$ decomposition steps for exactly representable weights, satisfying $W_T = 0$ for $T \leq F$. If early termination is applied after $T < F$ iterations, the remaining residual is bounded by 
\begin{equation}
|W_T| < 2^{-T}, 
\label{eq:dqmac_bound} 
\end{equation} so that the number of shift-and-add stages directly controls the approximation error, leading to gradual accuracy scaling. 

{
The proposed pre-accumulation truncation reduces the required accumulator width relative to conventional fixed-point multiplication. For two $N$-bit operands, a standard multiplier produces a $2N$-bit product, so the accumulator width for a dot product over $K$ terms is $2N + \lceil \log_2 K \rceil$ bits. The proposed DQ-MAC truncates each shifted partial product to $N$ bits prior to accumulation, reducing the accumulator width to $N + \lceil \log_2 K' \rceil$ bits, where $K'$ denotes the number of operands retained after pruning. Since $K' < K$, the accumulator width reduction is given by
}
\begin{equation}
\Delta B = B_{\text{conv}} - B_{\text{DQ-MAC}} = N + \lceil \log_2 K \rceil - \lceil \log_2 K' \rceil.
\label{eq:accum_reduction}
\end{equation}
For a $3{\times}3$ kernel, conventional accumulation requires $K=9$ terms, whereas the proposed pruning retains $K'=4$ operands, reducing the accumulator by $N+2$ bits. For $N=8$, the requirement decreases from 20 bits to 10 bits whereas for $N=4$, from 12 bits to 6 bits. This directly reduces adder width, registers, switching activity, and activation functions.

%\end{subequations}

\begin{figure}[!t]
    \centering
    \includegraphics[width=\columnwidth]{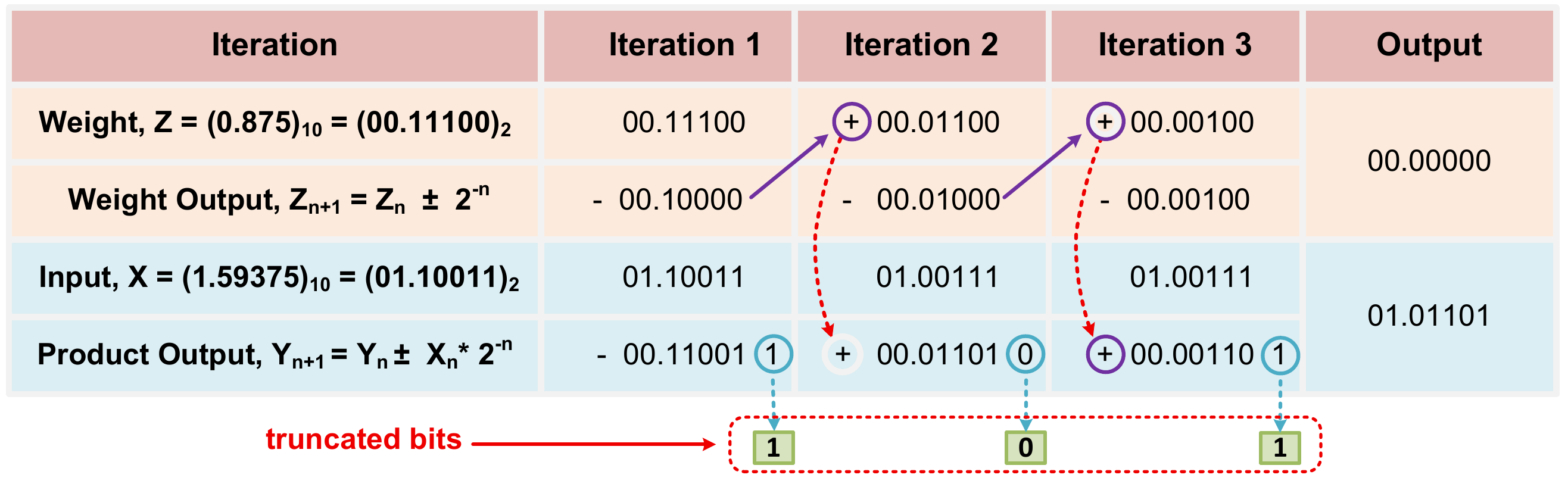}
    \caption{Example multiplication for the proposed algorithm with a
signed 8-bit fixed-point arithmetic scheme. Sampled calculation values
have been taken from one of the trained data samples.}
    \label{fig:algo_illustr}
\end{figure}

\subsection{Dual-Precision (4/8-bit) Quantised SIMD MAC (DQ-MAC) Unit for EV2AI acceleration}

\begin{figure}[!t]
    \centering
    \includegraphics[width=\columnwidth]{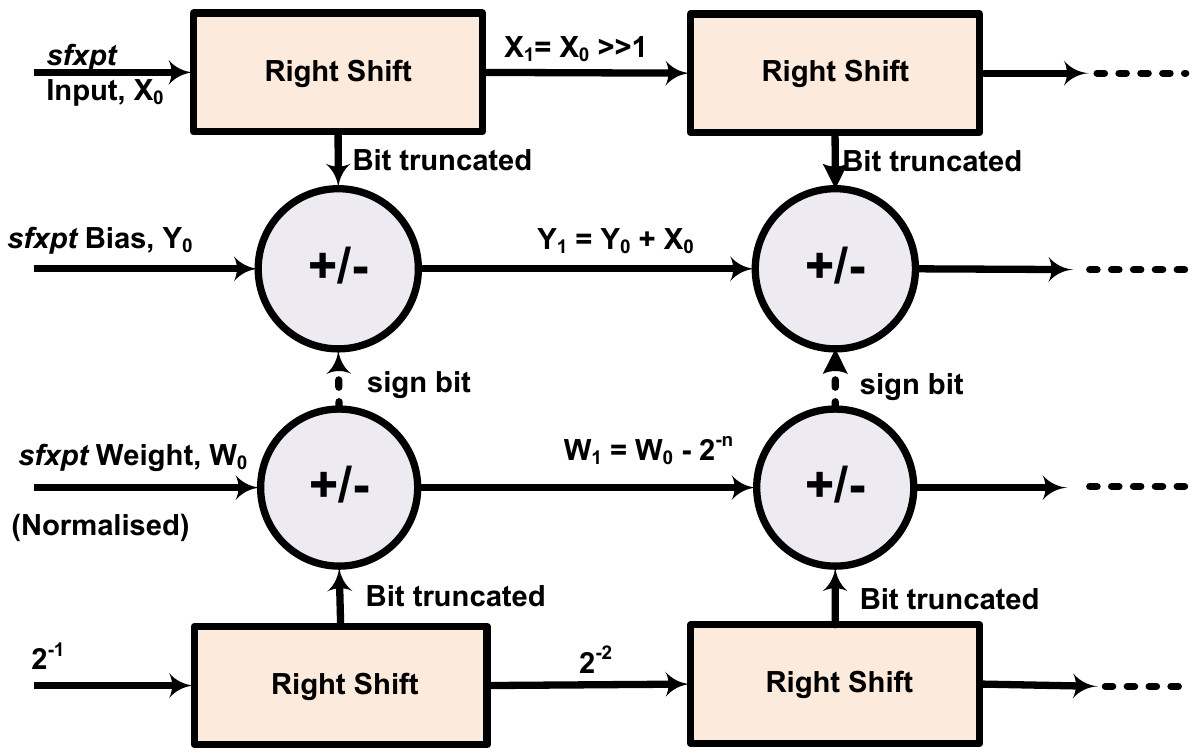}
    \caption{ Enhance performance pipeline architecture to increase
computational throughput. The design can be configured for both
single-stage iterative architecture and pipeline architecture to enhance
throughput}
    \label{fig:mac_compute_pipeline}
\end{figure}

The DQ-MAC unit supports both single-stage iterative and multi-stage pipelined configurations. In the iterative architecture, a single hardware instance is reused across $T$ clock cycles via a feedback path, conserving area and power at the cost of throughput. In the pipelined architecture with $P$ stages, dedicated stages operate concurrently to increase computational throughput, completing an MAC operation in fewer than $\lceil T/P \rceil$ clock cycles. As illustrated in Fig.~\ref{fig:mac_compute_pipeline}, each pipeline stage comprises a dedicated bit-shifter and an adder/subtractor block, with no multipliers employed. Three data signals propagate through the architecture: the feature map $\mathbf{X}_i$, the weight residual $\mathbf{W}_i$, and the accumulator output $\mathbf{Y}_i$, all updated at each stage according to Eq.~\eqref{eq:dqmac_output}.

\begin{figure}
    \centering
    \includegraphics[width=1\linewidth]{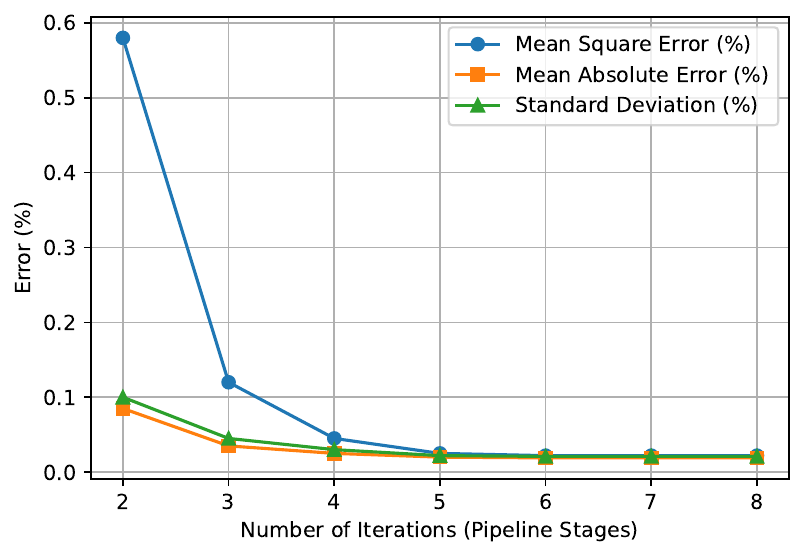}
    \caption{{The error evaluation distribution across various iterations for Pareto analysis and Pareto-point extraction.}}
    \label{fig:pareto}
\end{figure}

\begin{figure}[!t]
     \centering
     \includegraphics[width=0.95\columnwidth,height=42.5mm] {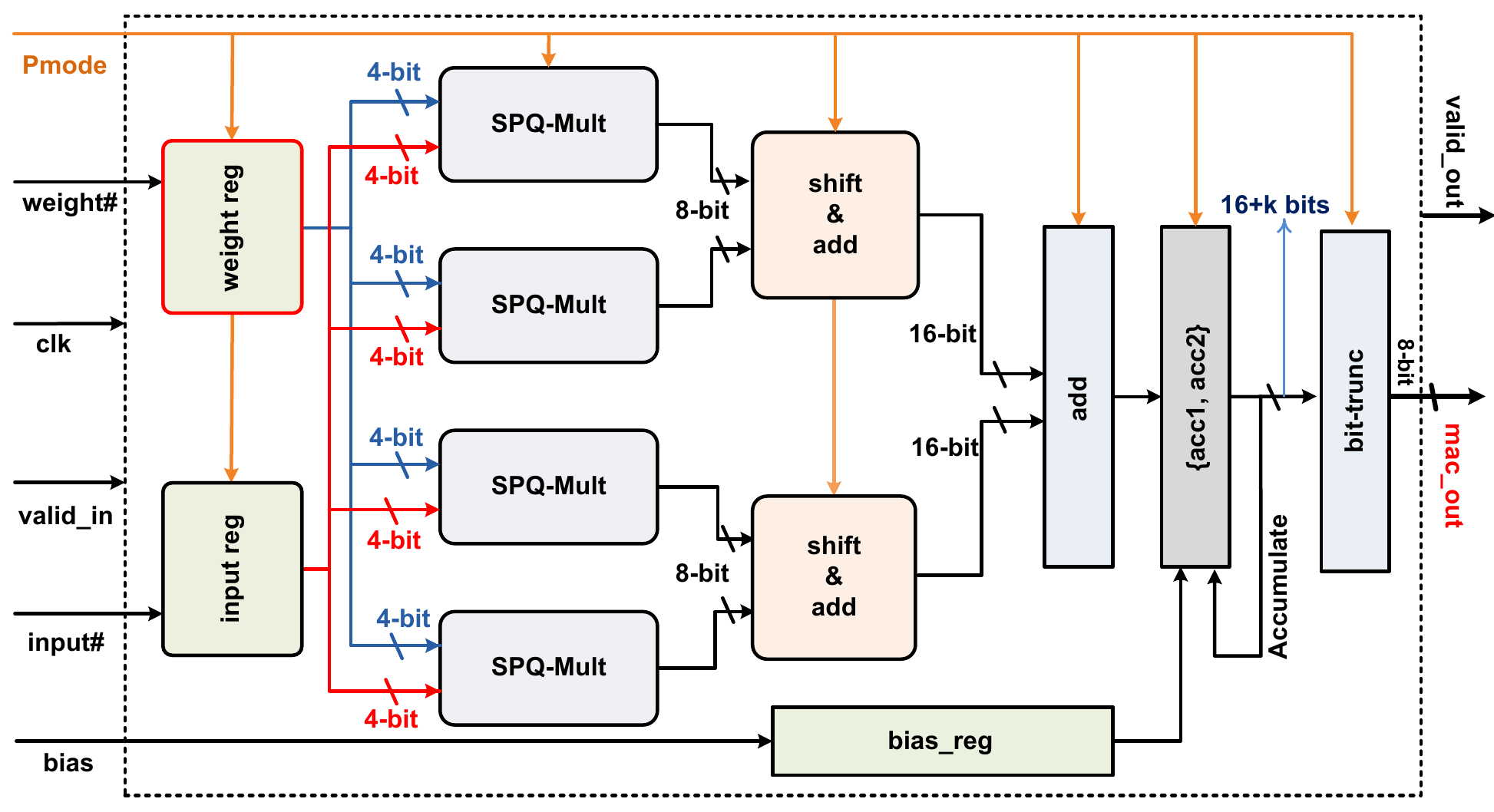}
     \caption {Enhanced SIMD DQ-MAC with 4/8-bit quantization reduces precision with 4x computational throughput.}
     \label{SIMD}    
\end{figure}

Fig.~\ref{SIMD} presents the top-level architecture of the proposed DQ-MAC unit, designed for SIMD-based processing of 4-bit and 8-bit fixed-point operands. At its core, the unit employs a single-precision Quant-Multiplier that leverages MSD-guided power-of-two quantisation in a pipelined manner. A Pareto analysis across varying pipeline depths (Fig.~\ref{fig:pareto}) shows that error metrics converge rapidly beyond five stages, establishing five pipeline stages as the design point that best balances accuracy, area, power, and critical-path delay. The single-precision unit is extended to a dual-precision 4/8-bit SIMD configuration through digit-level time-multiplexing of compute resources, achieving $4\times$ computational throughput without hardware duplication. Prior to inference, weights are normalised to maximum-normalised ($mn$) form and represented in a one-bit integer plus fractional fixed-point format.

 % \begin{figure}[!t]
 %     \centering
 %     \includegraphics[width=\columnwidth]{images/Fig_2_SIMD 48 fxp-MAC within Quant-PE.pdf}
 %     \caption{Integrated Dual-Quant-MAC with data flow and control signals, illustrating 1-output bit-widths.}
 %     \label{Quant-PE}    
 % \end{figure}

% Flowchart of the proposed algorithm for QuantMAC
% architecture. Here, the input feature map multiplies with the weights
% simply by shifting for each binary position and eliminating the weighted
% positioned bit in every iteration.

\subsection{Structured Hardware-Aware Reductive Pruning (SHARP) codesign strategy}

%Pruning directly addresses the latency and energy consumption challenges of EV2AI models by eliminating redundant parameters and, consequently, reducing the number of MAC operations. To this end, the proposed accelerator employs a static structured pruning strategy co-designed with the underlying MAC hardware architecture, termed Structured Hardware-Aware Reductive Pruning (SHARP).

% Ratko comment: TO_DO illustrate sharp procedure
%SHARP targets a 50\% pruning ratio, co-designed with the proposed MAC unit that computes a dot product over four 4-bit elements per cycle. For $3\times3$ kernels, SHARP applies 4:9 pruning by retaining the four largest weights by absolute value along with their indices, enabling the MAC unit to compute the full dot product in a single cycle with complete utilization. For $5\times5$ kernels, 12:25 pruning retains the twelve largest weights, completing the dot product in three fully utilized cycles. Retaining fewer elements would underutilize the MAC unit, while retaining more would introduce partially idle cycles, in both cases degrading hardware efficiency.

%\textcolor{red}{Structured pruning is performed using a static magnitude-based approach, where weights are ranked based on absolute magnitude and pruned prior to quantization-aware training (QAT). For hardware alignment, 4 weights are retained for 3×3 kernels and 12 weights for 5×5 kernels, ensuring full SIMD lane utilization. The pruning mask is fixed during QAT fine-tuning to co-optimize weight values with the proposed quantization scheme.}

To maximise hardware utilisation of the proposed SIMD DQ-MAC unit, a Structured Hardware-Aware Reductive Pruning (SHARP) strategy is introduced that jointly optimises sparsity, quantisation, and execution efficiency. Unlike conventional structured pruning, which primarily targets model compression, SHARP explicitly aligns pruning patterns with the SIMD execution granularity of the underlying hardware. 

The SHARP framework operates in three stages: (i) layer-wise precision assignment, (ii) structured pruning aligned to the SIMD width, and (iii) quantisation-aware fine-tuning. In the first stage, 4-bit quantisation is applied to all weights in a given layer. If the resulting accuracy degradation exceeds an acceptable threshold, the precision assignment for that layer is reverted to 8-bit.

In the second stage, SHARP enforces a structured sparsity pattern determined by the kernel size $N = k_h \times k_w$, retaining
\begin{equation}
R = 4 \cdot \left\lfloor N/8 \right\rfloor
\end{equation}
weights per kernel, where $R$ is always a multiple of four and thus aligned with the four-lane SIMD datapath of the DQ-MAC unit. For $3{\times}3$ kernels ($N=9$), this yields $R=4$ (4:9 pruning ratio), enabling single-cycle execution in 4-bit mode. For $5{\times}5$ kernels ($N=25$), $R=12$ (12:25 pruning ratio), requiring three execution cycles. In both cases, the reduced operand count directly lowers accumulator pressure. By co-designing structured sparsity with the MAC hardware and quantisation scheme, SHARP simultaneously reduces memory access latency, energy consumption, and total MAC operation count, while guaranteeing full SIMD lane utilisation and eliminating partially occupied execution cycles.

In the third stage, quantisation-aware fine-tuning is applied with a fixed sparsity mask, ensuring that retained weight values are co-optimised with the DQ-MAC's power-of-two approximation scheme. Since the retained weights are guaranteed to be a multiple of the SIMD width, the resulting execution schedule achieves full lane utilisation per cycle, reduced execution latency, lower memory access overhead, and elimination of idle compute cycles. Consequently, SHARP serves as both a model compression strategy and a hardware scheduling mechanism.

\begin{table}[t]
\centering
\caption{Ablation analysis on YOLO-based EV2AI inference.}
\label{tab:sharp_ablation}
\resizebox{\columnwidth}{!}{
\begin{tabular}{lcccccc}
\hline
\textbf{Configuration} &
\textbf{Precision} &
\textbf{Pruning} &
\textbf{SIMD Util.} &
\textbf{Kernel Cycles} &
\textbf{mAP Drop} &
\textbf{Latency Gain} \\
\hline
FP32 baseline &
FP32 &
None &
- &
9 / 25 &
0\% &
1.0$\times$ \\

Uniform 8-bit &
8-bit &
None &
100\% &
9 / 25 &
$<$1\% &
1.0$\times$ \\

AHCO~\cite{AHCO-YOLO} &
4/8-bit &
- &
70-85\% &
3-4 / 8-10 &
1-3\% &
2.5-3.5$\times$ \\

Uniform 4-bit &
4-bit &
None &
100\% &
3 / 7 &
2-4\% &
3.0-3.6$\times$ \\

Bose-8\cite{POLARON} &
FP-4/ Posit-8 &
6:9 &
82.75\% &
5 / 13 &
1.25\% &
3.4-4.9$\times$ \\

Structured pruning &
8-bit &
50\% unaligned &
50-75\% &
5 / 13 &
1-2\% &
1.8-1.9$\times$ \\

DQ-MAC only &
4/8-bit &
None &
100\% &
3 / 7 &
$<$3\% &
3.0-3.6$\times$ \\

SHARP only &
8-bit &
4:9 / 12:25 &
100\% &
4 / 12 &
1-2\% &
2.1-2.25$\times$ \\

\textbf{DQ-MAC + SHARP} &
\textbf{4/8-bit} &
\textbf{4:9 / 12:25} &
\textbf{100\%} &
\textbf{1 / 3} &
\textbf{$<$3\%} &
\textbf{8.3-9.0$\times$} \\
\hline
\end{tabular}}
\end{table}

Table~\ref{tab:sharp_ablation} summarises the contribution of each component in the proposed SHARP-DQ-MAC co-design. Uniform 4-bit quantisation improves latency but may introduce accuracy degradation in sensitive layers. Standard structured pruning reduces the number of active weights, but its retained weights are not necessarily aligned with the four-lane SIMD datapath, leading to partially occupied execution cycles. In contrast, SHARP retains 4 and 12 weights for $3 \times 3$ and $5 \times 5$ kernels, respectively, ensuring full SIMD utilization. When combined with the dual-precision DQ-MAC, the proposed approach reduces the execution of $3 \times 3$ kernels from 9 cycles to 1 cycle and $5 \times 5$ kernels from 25 cycles to 3 cycles, while maintaining less than 3\% mAP degradation after quantisation-aware fine-tuning.

\subsection{Reconfigurable Quantise-enabled (RQ-NAF) non-linear activation function Core}

\begin{figure}[!t]
     \centering
      \includegraphics[width=85mm,height=40mm] {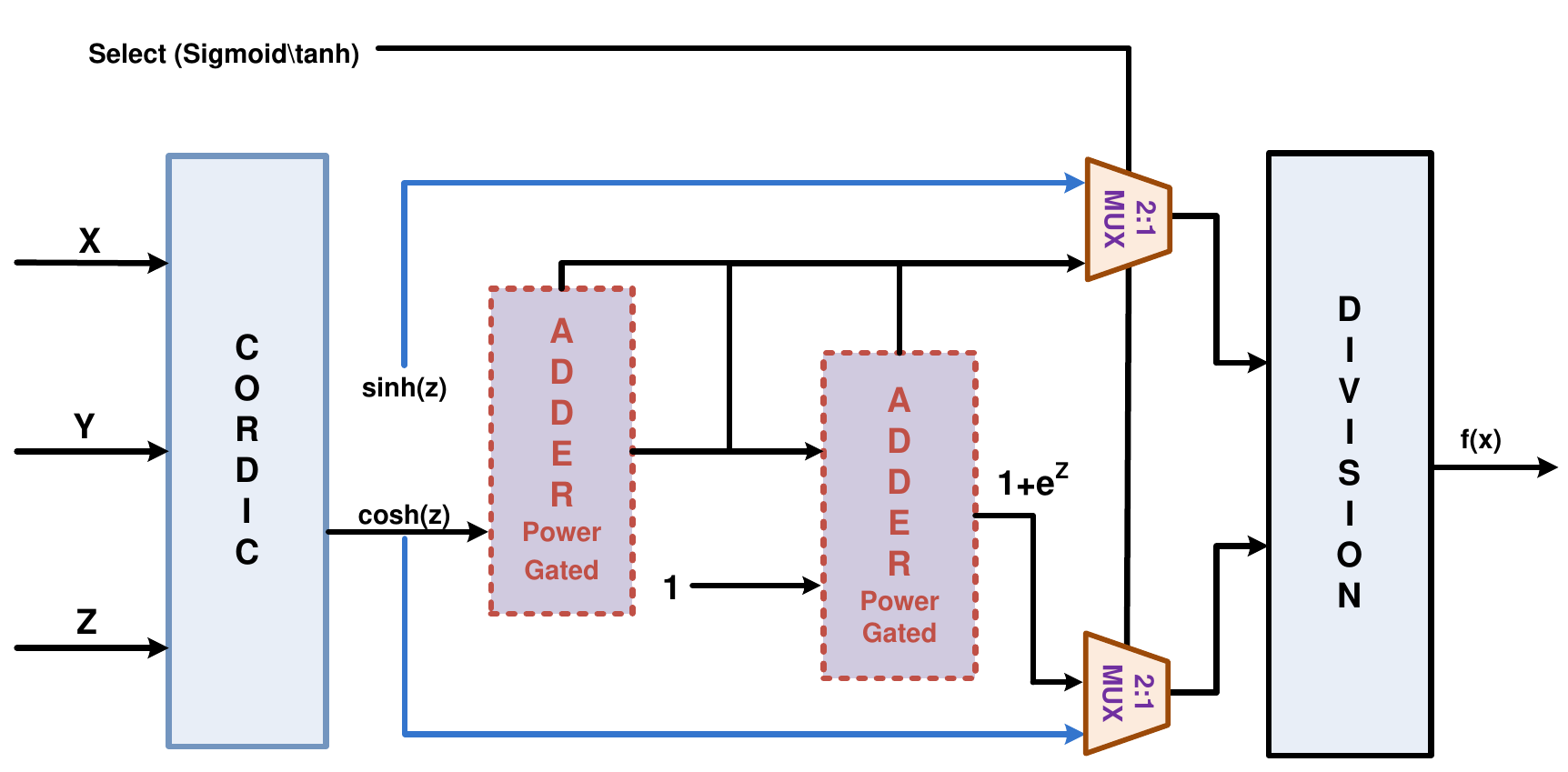}
     \caption{Block level of architecture showing data flow for runtime-configurable CORDIC-based activation function beyond ReLU}
    % \vspace{-6.5mm}
     \label{Config-AF}
\end{figure}

RQ-NAF core is implemented as a reconfigurable CORDIC-based datapath integrated within the processing element, as shown in Fig. \ref{Config-AF}. The activation engine employs a 9-stage pipelined CORDIC architecture that, following an initial pipeline fill latency of 9 cycles, produces one activation output per cycle at full throughput. A 2-bit control signal enables runtime selection among ReLU, Sigmoid and Tang operations over a shared arithmetic datapath, eliminating the need to replicate activation hardware across supported network topologies. The fixed-point CORDIC implementation is ultra-low power and area-efficient, supporting CNNs and Transformer MLP blocks within a single instantiation while maintaining low approximation error relative to floating-point evaluation.

Since EV2AI workloads require significantly fewer activation evaluations than MAC operations, a single RQ-NAF instance is time-multiplexed, avoiding the area overhead of per-layer activation hardware duplication. Combined with the MSD-PoT quantization scheme and SHARP pruning, this reconfigurable reuse strategy collectively optimises resource efficiency, positioning the proposed accelerator as a practical solution for area- and power-constrained edge deployment.

\subsection{Time-Multiplexed, Resource-Efficient Edge Accelerator for Object Detection and Classification}

\begin{figure}[!t]
     \centering
    \includegraphics[scale=0.3] {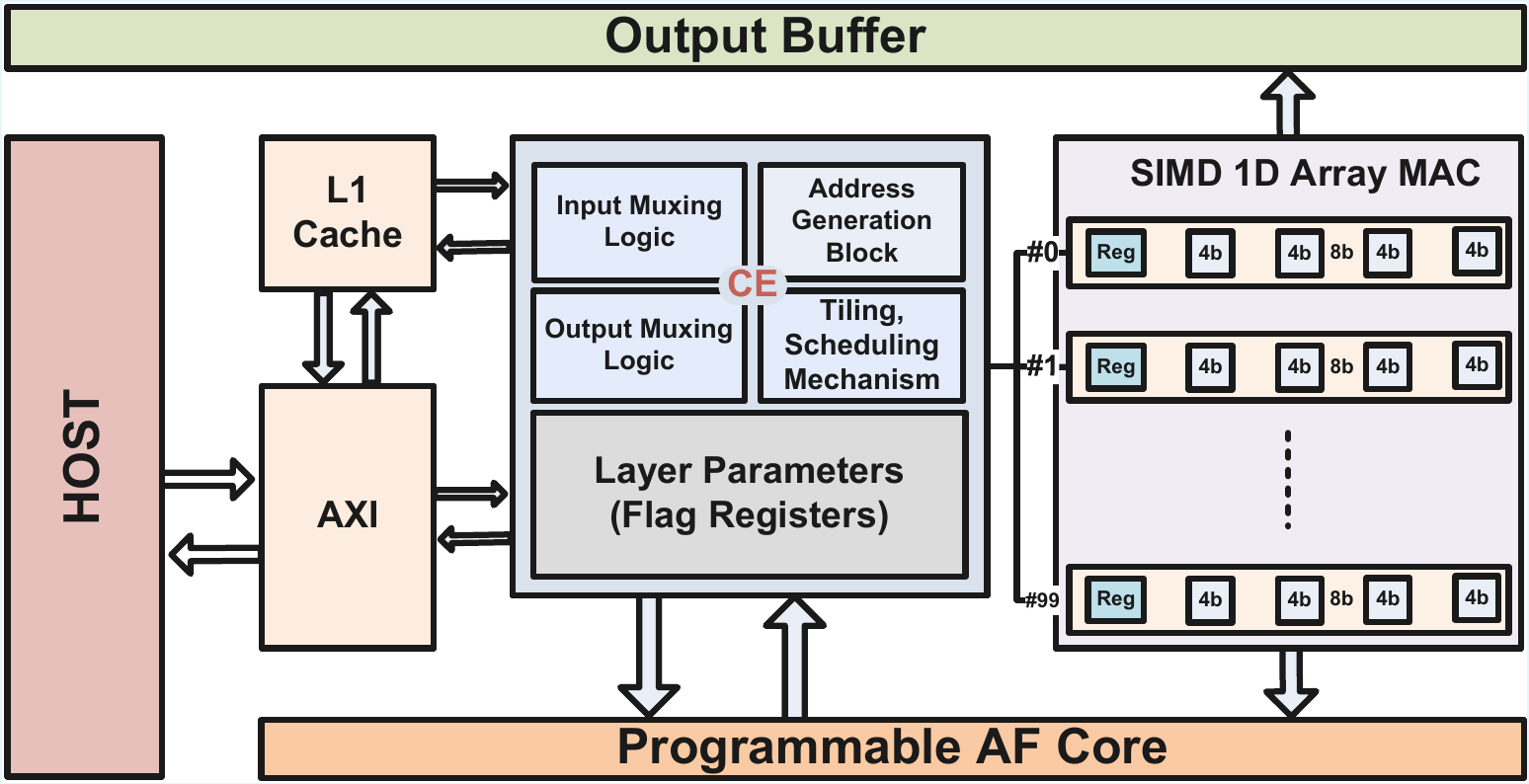}
    \caption{Micro-architecture of Time-Multiplexed Resource-Efficient Edge-AI Accelerator with control engine and Host interfacing.}
     %\vspace{-6.5mm}
     \label{TREA}
\end{figure}

The TREA accelerator is organised as a 1D array of 100 SIMD 4-bit DQ-MAC units, as shown in Fig. \ref{TREA}. Each DQ-MAC unit serves as the primary compute primitive, performing fixed-point MAC operations between filter maps and activation inputs, with bias values preloaded into the accumulator before the MAC computation phase to eliminate the need for dedicated cycles and reduce overall latency. The design adopts a time-multiplexed single-layer hardware structure, dynamically reconfigured to support varying DNN sizes and topologies without replicating dedicated hardware per layer. This parameterised architecture minimises intra-layer data movement dependencies, simplifies the control engine, and enables linear scalability across both FPGA and ASIC targets.

Resource efficiency is further enhanced through activation function hardware reuse via a Parallel-In Serial-Out (PISO) architecture, wherein a single RQ-NAF is shared across all neurons within a layer rather than instantiated per neuron, reducing per-neuron area overhead at the cost of marginally increased latency. On-chip weight and bias storage is implemented using BRAM, while temporary intermediate data is managed through FIFO buffers. Layer-wise execution is coordinated through three principal control signals: \texttt{Compute\_Done} triggers processing of the subsequent tile upon completion, \texttt{Layer\_Done} signals the end of a fully-connected or convolutional layer, and \texttt{DNN\_Done} communicates full workload completion to the host processor. This time-multiplexed AF reuse strategy reduces DNN area occupation by approximately one order of magnitude, achieving an effective balance between throughput and chip area for resource-constrained edge EV2AI deployment.

\begin{figure*}
    \centering
    \includegraphics[width=1\textwidth]{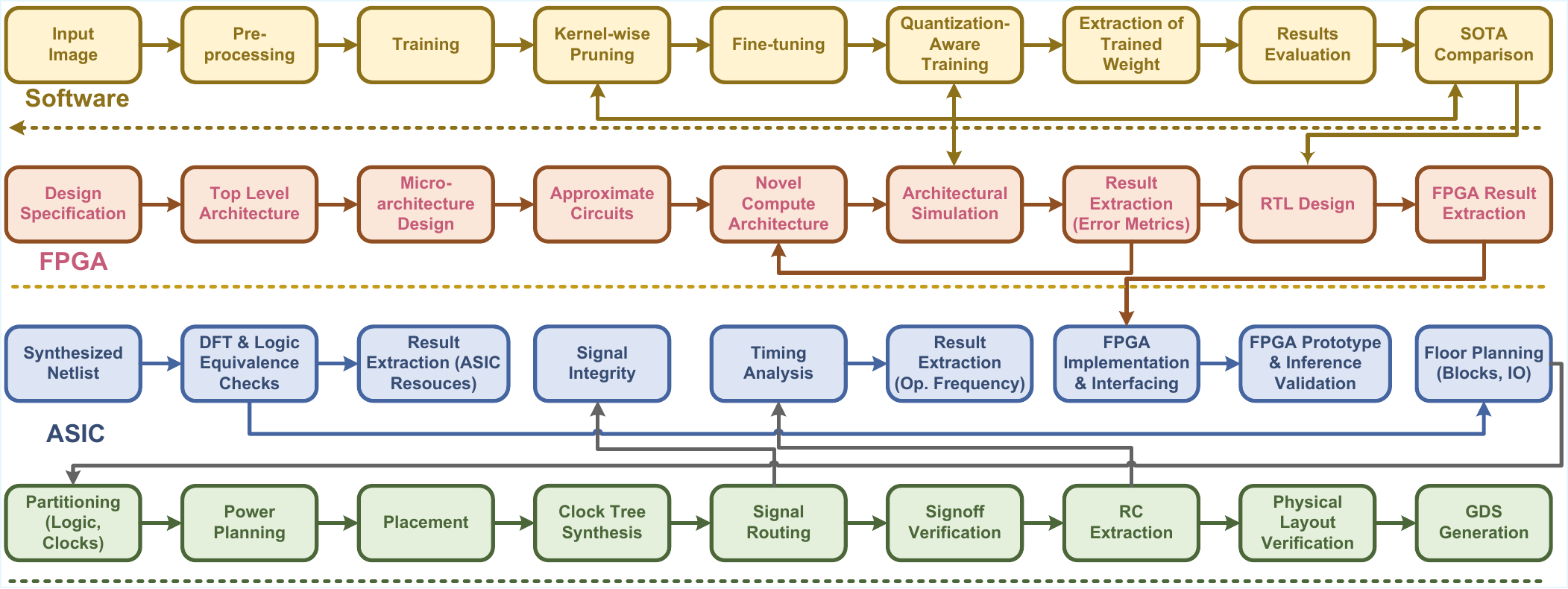}
    \caption{Detailed methodology/workflow illustrating Algorithm–Hardware co-design followed for the empirical evaluation, incorporating Dual-precision Quantization and SHARP.}
    \label{fig:methodology}
\end{figure*}

\section{Methodology \& Performance Evaluation}

Fig.~\ref{fig:methodology} illustrates the evaluation workflow for the proposed design, encompassing three complementary domains: software simulation, FPGA prototyping, and ASIC implementation. In the software domain, Python-based models provide iso-functional emulation of the proposed architecture; all experiments were executed on an NVIDIA L4 GPU within a Google Colab environment. For hardware evaluation, the design was described in Verilog HDL and functionally verified using QuestaSim, with cross-validation performed against the software emulation framework. FPGA evaluation was carried out using the AMD Vivado Design Suite, targeting the Xilinx VC707 platform, from which post-implementation resource utilisation metrics were extracted. ASIC evaluation was performed using the Cadence RTL-to-GDS flow, employing Genus for logic synthesis and Innovus for physical implementation, targeting a CMOS HPC 28-nm process node, with post-route timing, area, and power metrics subsequently reported.

\subsection{{Training setup}}
For object detection, YOLO variants (Tiny, Small, Medium, and Large) are evaluated on PASCAL VOC 2007+2012 and COCO 2017 using standard splits. All models are trained in FP32 at $416{\times}416$ resolution for 200--300 epochs with SGD (momentum 0.9, weight decay $5{\times}10^{-4}$), an initial learning rate of 0.01, and cosine or step decay. Augmentation includes mosaic, random flipping, scaling, and colour jittering. For image classification, MobileNet, EfficientNet, and Inception models are trained on ImageNet (ILSVRC-2012) for 90 epochs at $224{\times}224$ resolution with SGD (momentum 0.9, weight decay $1{\times}10^{-4}$), an initial learning rate of 0.1, and step decay at epochs 30, 60, and 80, with random cropping, horizontal flipping, and normalisation applied throughout. All models are first trained in FP32, then subjected to SHARP structured pruning followed by quantization-aware fine-tuning (QAT). SHARP adopts a static magnitude-based criterion, retaining four weights per $3{\times}3$ kernel and twelve per $5{\times}5$ kernel to enforce hardware-aligned sparsity. The pruning mask is fixed during QAT, and dual-precision 4/8-bit quantization is applied layer-wise to jointly optimize accuracy and hardware efficiency.

\subsection{Impact on classification accuracy}

\begin{figure}
    \centering
    \includegraphics[width=\columnwidth]{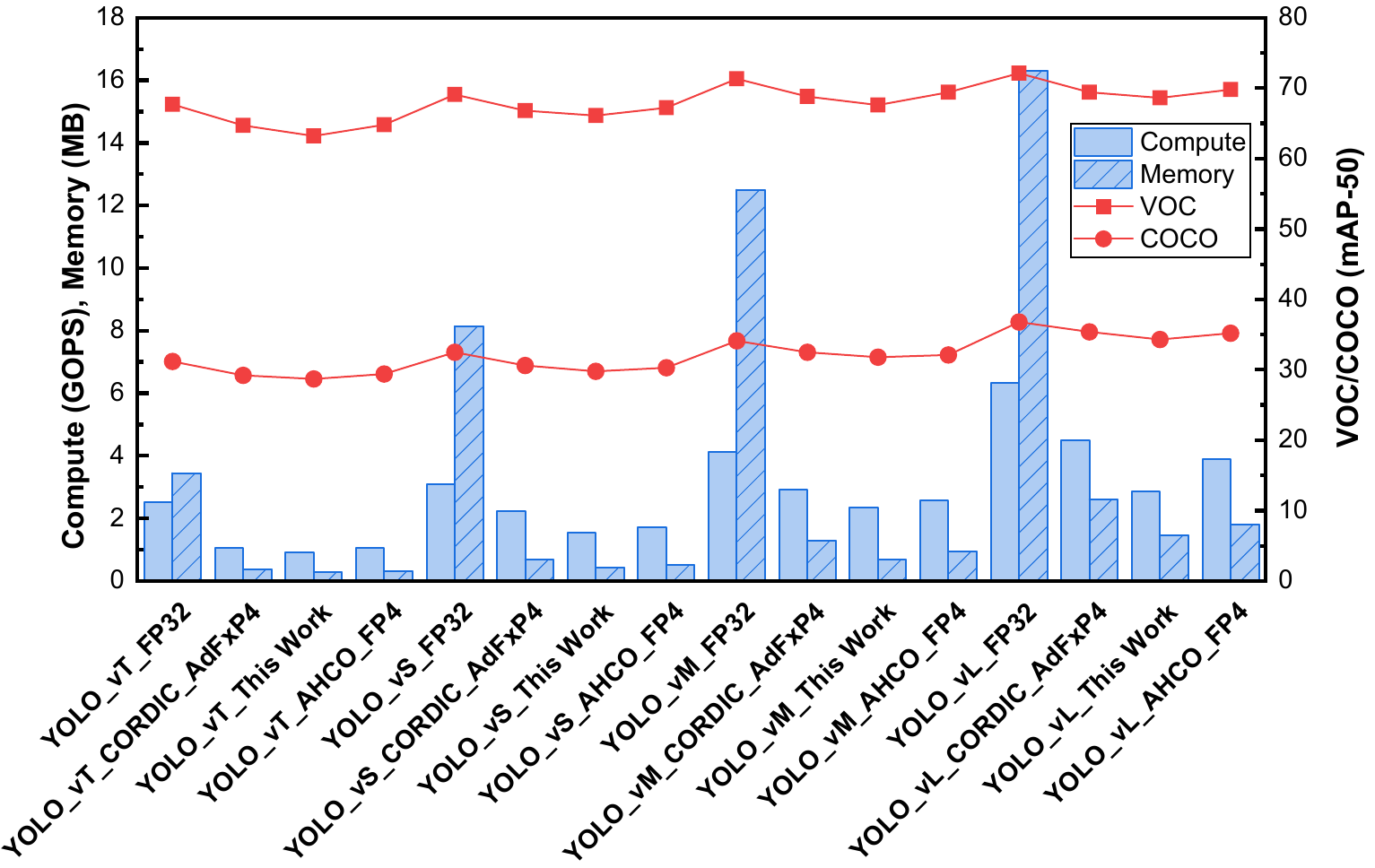}
    \caption{Software Performance comparison baseline (FP32) and (4-bit quantization AdFxP, DQ, and FP4) with \cite{Flex-PE} and \cite{AHCO-YOLO}.}
    \label{fig:comp_yolo_4b}
\end{figure}

We evaluate the DQ-MAC unit on object detection within the EV2AI system, targeting resource-constrained edge devices. Several scalable YOLO variants~\cite{AHCO-YOLO} (v\_T- 1.9 M, v\_S 7.2M, v\_M 21.2M and v\_L 46.5 parameters) were trained at FP32 precision and quantized using either the proposed DQ-MAC, with dynamic per-layer bit-width assignment (4- or 8-bit) or fixed bit-width state-of-the-art MAC designs~\cite{Flex-PE,AHCO-YOLO}. Detection performance (mAP@0.5, memory footprint, and GOPS) was assessed on PASCAL VOC~\cite{Everingham2009ThePV} and COCO~\cite{yi2014COCO}, with results shown in Fig.~\ref{fig:comp_yolo_4b}.

The proposed DQ-MAC unit achieves detection accuracy comparable to the FP32 baseline across all evaluated YOLO variants on both datasets, with less than 3\% mAP@0.5 degradation. The DQ-MAC consistently outperforms state-of-the-art fixed-bit-width MAC designs, particularly at 4-bit precision, while reducing memory footprint and computational demand. These gains are most pronounced for larger YOLO variants (Medium and Large), demonstrating that dynamic per-layer quantisation effectively balances accuracy and resource efficiency for edge deployment.

\begin{figure}
    \centering
    \includegraphics[width=\columnwidth]{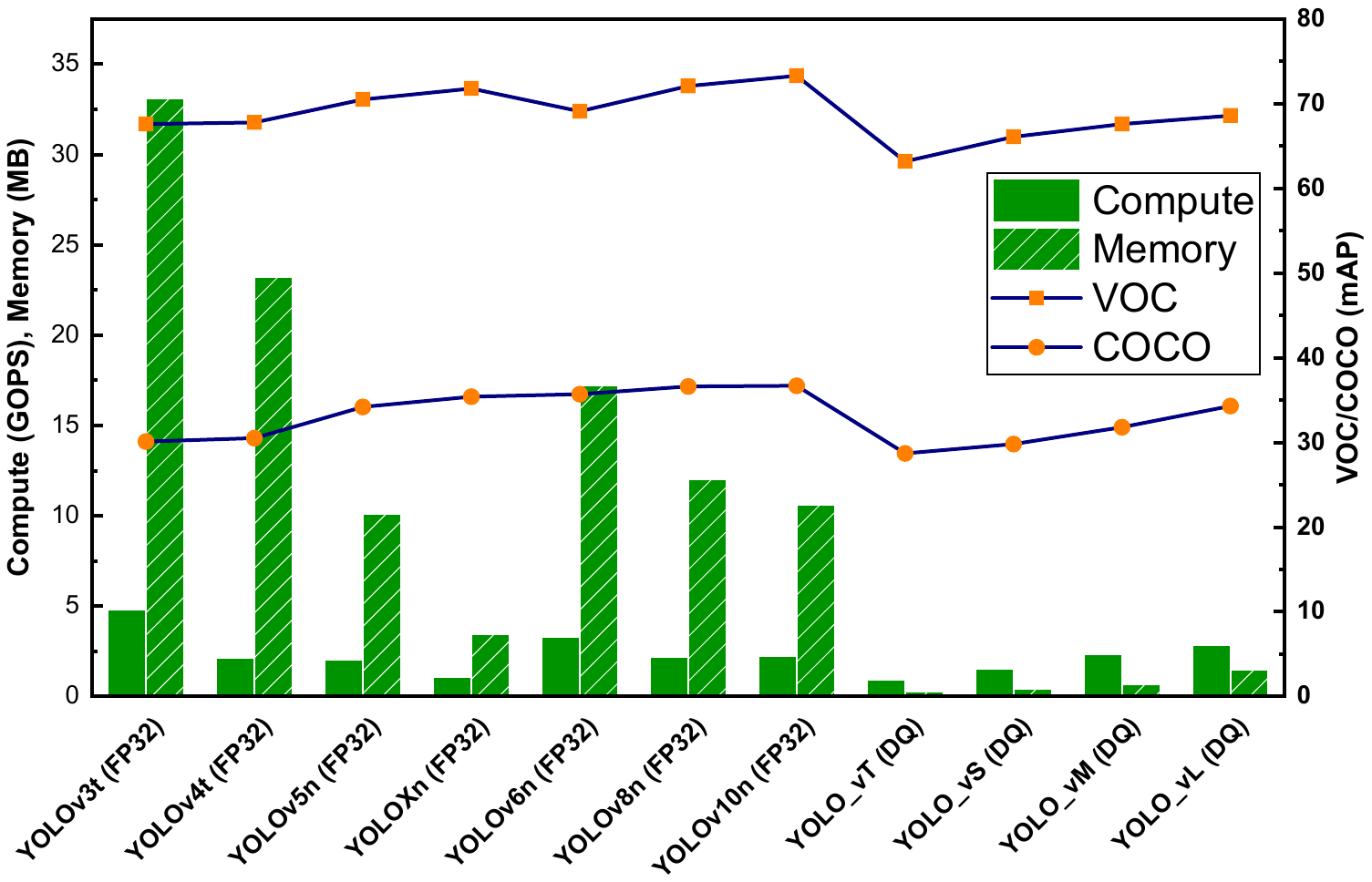}
    \caption{Software Performance comparison for different generations of object detection (YOLO) models, illustrating model parameters (memory MB), Compute ops. reqd. (GOPS), mAP-50 for VOC and COCO.}
    \label{fig:yolo_comp}
\end{figure}

Fig.~\ref{fig:yolo_comp} further compares DQ-MAC-equipped YOLO variants against more recent YOLO generations. The DQ-MAC scales favourably with model complexity, with mAP@0.5 increasing from Tiny to Large variants while model size and GOPS remain substantially lower. The YOLO Large variant with DQ-MAC achieves accuracy comparable to YOLOv8 and YOLOv10 on both PASCAL VOC and COCO, with significantly fewer parameters and operations, confirming the efficacy of the proposed design for accurate, resource-efficient edge inference.

\begin{figure}[!t]
    \centering
    \includegraphics[width=\columnwidth]{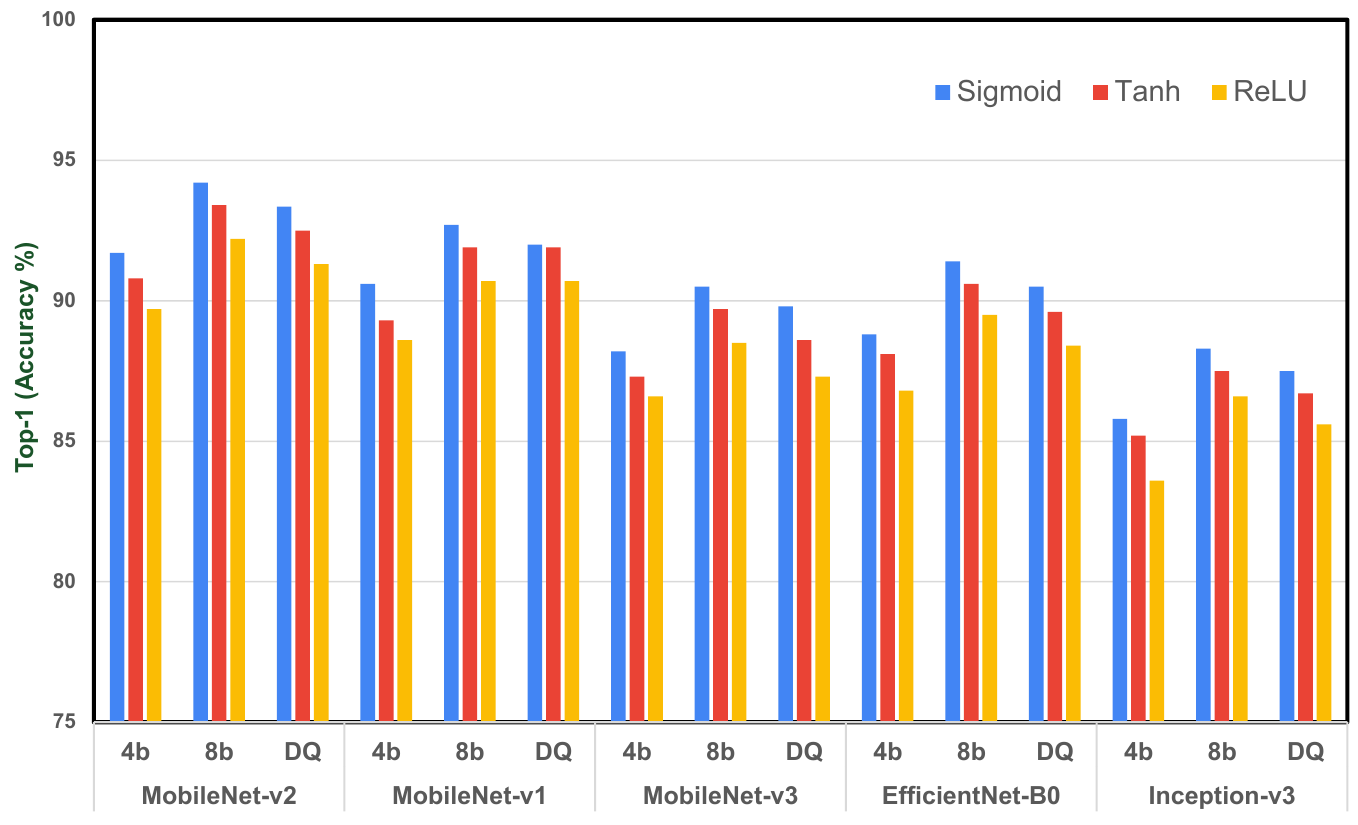}
    \caption{Implications of RQ-NAF on AI Workloads}
    \label{fig:af_acc}
\end{figure}

The impact of the proposed RQ-NAF core on classification accuracy is evaluated on MobileNet~\cite{mobilenets}, EfficientNet~\cite{tan2019efficientnet}, and Inception~\cite{szegedy2015going} architectures trained on ImageNet~\cite{ImageNet}. The RQ-NAF core supports runtime-selectable activation functions (ReLU, Sigmoid, and Tanh), assessed under three quantisation schemes: 4-bit uniform, 8-bit uniform, and the proposed dual-precision configuration. Fig.~\ref{fig:af_acc} reports top-1 accuracy across all combinations. The core maintains stable accuracy across all activation functions and quantisation, with dual-precision achieving accuracy comparable to 8-bit uniform quantisation while selectively operating at 4-bit precision in designated layers. These results confirm the suitability of the RQ-NAF core for efficient dual-precision inference within the EV2AI system.

\subsection{Evaluation of DQ-MAC unit}

The hardware efficiency of the proposed DQ-MAC unit is evaluated on both FPGA and ASIC platforms, assessing its suitability for multiply-accumulate (MAC) operations across a range of design objectives. Table~\ref{tab:FPGA-scale} presents scalability results with respect to operand bit-width, alongside the Xilinx IP core~\cite{Xilinx-IP} as a reference. The proposed DQ-MAC unit achieves reductions in LUT utilisation of up to 72\% and power consumption of up to 50\% relative to the reference design, both critical for energy-constrained EV2AI deployments. The reduced critical-path delay is particularly advantageous for latency-sensitive applications such as real-time object detection. Collectively, these results demonstrate that the proposed design scales efficiently with increasing operand bit-width, retaining its efficiency advantages at higher precisions.

\begin{table}[!t]
\caption{Quantifying the effect with precision scalability on MAC FPGA resources, compared with Xilinx IP\cite{Xilinx-IP}.}
\label{tab:FPGA-scale}
\renewcommand{\arraystretch}{1.35}
\resizebox{\columnwidth}{!}{%
\begin{tabular}{|l|cccc|llll|}
\hline
\multirow{2}{*}{\textbf{Metrics}} & \multicolumn{4}{c|}{\textit{\textbf{Xilinx IP \cite{Xilinx-IP} }}} & \multicolumn{4}{c|}{\textit{\textbf{This Work \color[HTML]{009901}(Implications, same bit-precision) }}} \\ \cline{2-9} 
 & \multicolumn{1}{c|}{\textbf{4-bit}} & \multicolumn{1}{c|}{\textbf{8-bit}} & \multicolumn{1}{c|}{\textbf{16-bit}} & \textbf{32-bit} & \multicolumn{1}{c|}{\textbf{4-bit}} & \multicolumn{1}{c|}{\textbf{8-bit}} & \multicolumn{1}{c|}{\textbf{16-bit}} & \textbf{32-bit} \\ \hline
 
\textbf{LUTs} & \multicolumn{1}{c|}{37} & \multicolumn{1}{c|}{88} & \multicolumn{1}{c|}{334} & 1216 & \multicolumn{1}{c|}{\begin{tabular}[c]{@{}c@{}}24 \\ \color[HTML]{009901}(-35\%)\end{tabular}} & \multicolumn{1}{c|}{\begin{tabular}[c]{@{}c@{}}52 \\ \color[HTML]{009901}(-41\%)\end{tabular}} & \multicolumn{1}{c|}{\begin{tabular}[c]{@{}c@{}}106 \\ \color[HTML]{009901}(-68.3\%)\end{tabular}} & \begin{tabular}[c]{@{}c@{}}335 \\ \color[HTML]{009901}(-72.5\%)\end{tabular} \\ \hline

\textbf{FFs} & \multicolumn{1}{c|}{28} & \multicolumn{1}{c|}{76} & \multicolumn{1}{c|}{134} & 212 & \multicolumn{1}{c|}{\begin{tabular}[c]{@{}c@{}}16 \\ \color[HTML]{009901}(-43\%)\end{tabular}} & \multicolumn{1}{c|}{\begin{tabular}[c]{@{}c@{}}88 \\ \color[HTML]{FE0000}(+15.8\%)\end{tabular}} & \multicolumn{1}{c|}{\begin{tabular}[c]{@{}c@{}}168 \\ \color[HTML]{FE0000}(+25.4\%)\end{tabular}} & \begin{tabular}[c]{@{}c@{}}252 \\ \color[HTML]{FE0000}(+19\%)\end{tabular} \\ \hline

\textbf{CPD (ns)} & \multicolumn{1}{c|}{2.18} & \multicolumn{1}{c|}{2.37} & \multicolumn{1}{c|}{2.93} & 6.13 & \multicolumn{1}{c|}{\begin{tabular}[c]{@{}c@{}}1.01 \\ \color[HTML]{009901}(-53.7\%)\end{tabular}} & \multicolumn{1}{c|}{\begin{tabular}[c]{@{}c@{}}1.57 \\ \color[HTML]{009901}(-33.7\%)\end{tabular}} & \multicolumn{1}{c|}{\begin{tabular}[c]{@{}c@{}}2.21 \\ \color[HTML]{009901}(-24.6\%)\end{tabular}} & \begin{tabular}[c]{@{}c@{}}3.51 \\ \color[HTML]{009901}(-42.7\%)\end{tabular} \\ \hline

\textbf{Power (mW)} & \multicolumn{1}{c|}{2.2} & \multicolumn{1}{c|}{6.6} & \multicolumn{1}{c|}{14.8} & 24 & \multicolumn{1}{c|}{2.23}& \multicolumn{1}{c|}{\begin{tabular}[c]{@{}c@{}}6.36 \\ \color[HTML]{009901}(-3.6\%)\end{tabular}} & \multicolumn{1}{c|}{\begin{tabular}[c]{@{}c@{}}11.77 \\ \color[HTML]{009901}(-20.5\%)\end{tabular}} & \begin{tabular}[c]{@{}c@{}}21.78 \\ \color[HTML]{009901}(-9.3\%)\end{tabular} \\ \hline

\textbf{PDP (pJ)} & \multicolumn{1}{c|}{4.78} & \multicolumn{1}{c|}{15.64} & \multicolumn{1}{c|}{43.36} & 147 & \multicolumn{1}{c|}{\begin{tabular}[c]{@{}c@{}}2.3 \\ \color[HTML]{009901}(-52\%)\end{tabular}} & \multicolumn{1}{c|}{\begin{tabular}[c]{@{}c@{}}10 \\ \color[HTML]{009901}(-36\%)\end{tabular}} & \multicolumn{1}{c|}{\begin{tabular}[c]{@{}c@{}}26 \\ \color[HTML]{009901}(-40\%)\end{tabular}} & \begin{tabular}[c]{@{}c@{}}76.5 \\ \color[HTML]{009901}(-48\%)\end{tabular} \\ \hline

\end{tabular}}
\end{table}

\begin{figure}[!t]
    \centering
    \includegraphics[width=\columnwidth]{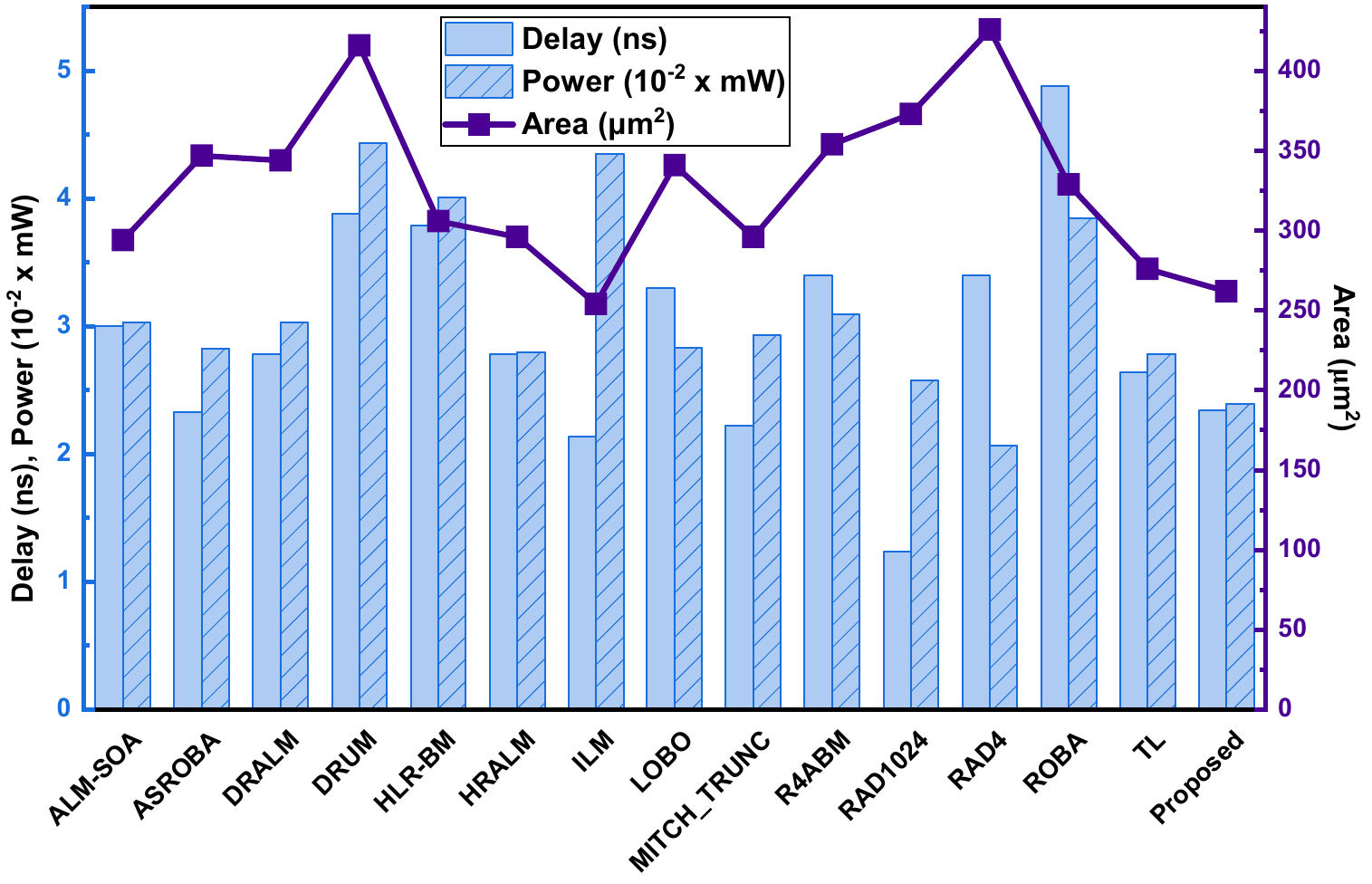}
    \caption{ASIC resource-comparison with prior approximate MAC units, data adapted from \cite{TL}.}
    %\cite{TL, DR-ALM, ILM-AA, MITCH_TRUNC, MITCH, ALM_SOA, AS-ROBA, DRUM, R4ABM, RAD1024, HLR-BM, LOBO, Posit-ILM, SS-MAC,MSDF-MAC_TVLSI'25}.}
    \label{fig:asic_app8_mac}
\end{figure}

The proposed DQ-MAC unit is further benchmarked against state-of-the-art approximate MAC designs in terms of area, power consumption, delay, energy efficiency, and power-delay product (PDP), with ASIC results from a CMOS 28-nm standard-cell library presented in Fig.~\ref{fig:asic_app8_mac}. The proposed design demonstrates competitive area, delay, and energy consumption across all evaluated implementations. While certain designs, such as RAD1024~\cite{RAD1024} and ILM~\cite{ILM-AA}, achieve lower delay or power consumption in isolation, the proposed DQ-MAC offers a more compact and balanced implementation, better suited for area- and power-constrained embedded systems. These characteristics stem from the co-designed MAC architecture, which tightly integrates shift-and-add operations with MSDF-based processing. Unlike prior works, where multiplier and adder stages are typically optimised independently, the proposed design jointly optimises both stages into a unified datapath, yielding reductions in critical-path delay and improved energy efficiency.

\begin{figure}[!t]
    \centering
    \includegraphics[width=\columnwidth]{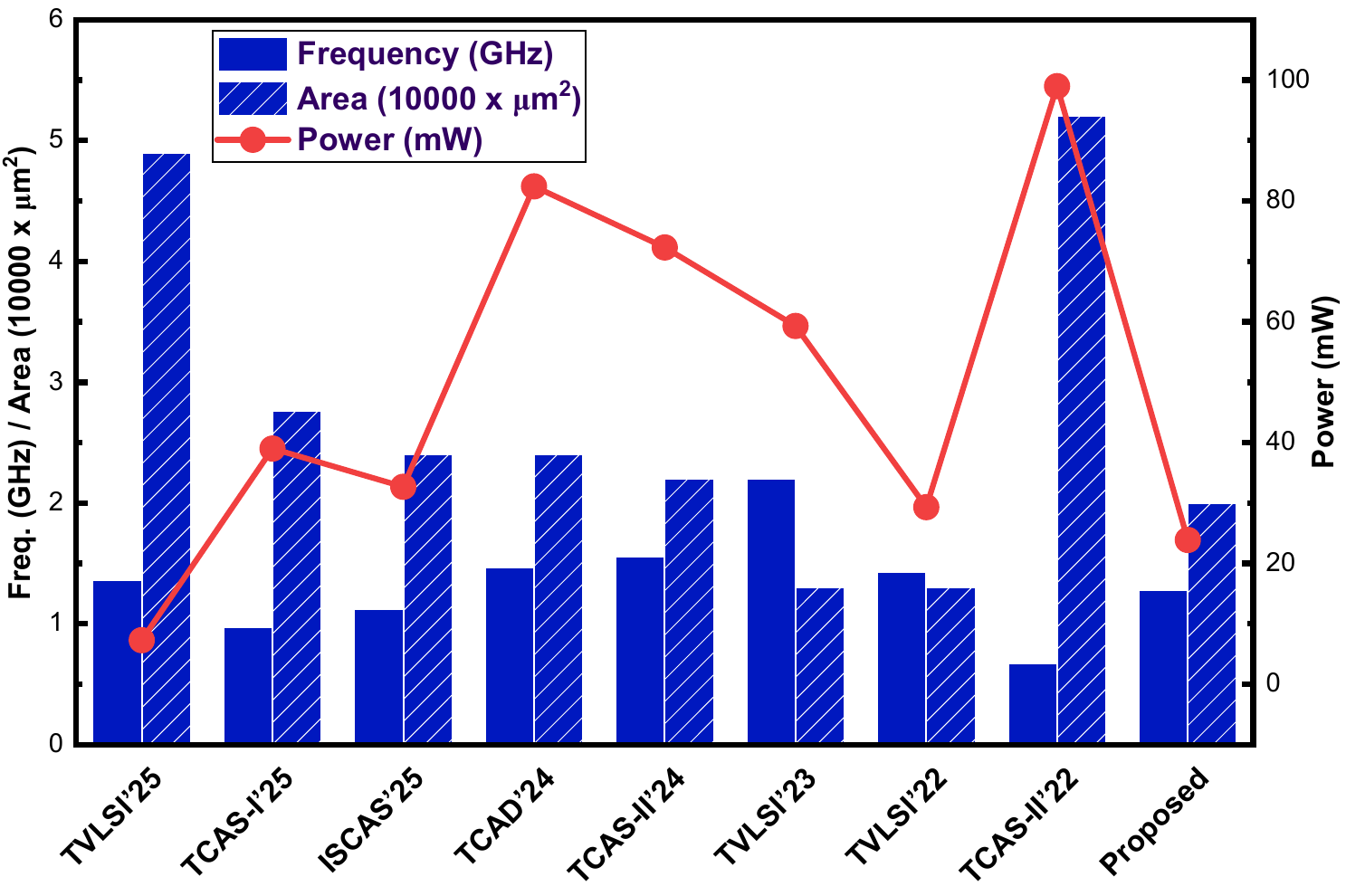}
    \caption{ASIC resource-comparison with prior SIMD MAC units\cite{Flex-PE, Maestro, LPRE, DPDAC-TCAD'24, RPE_TCAS-II'24, M-FMA, FP-MPE, UVMAC-TCASII'22}.}
    \label{fig:asic_simd_mac}
\end{figure}

The proposed DQ-MAC unit is benchmarked as a SIMD processing element against state-of-the-art neural network accelerator MAC units, with ASIC results in Fig.~\ref{fig:asic_simd_mac}. The DQ-MAC achieves frequency comparable to most evaluated designs; the exception is~\cite{UVMAC-TCASII'22}, whose deeply pipelined floating-point architecture yields higher frequency at the cost of increased area and power. In terms of area, the DQ-MAC outperforms~\cite{Flex-PE, Maestro, LPRE, DPDAC-TCAD'24, RPE_TCAS-II'24, UVMAC-TCASII'22} and represents the most compact design among those supporting dynamic bit-width reconfigurability, though slightly exceeding~\cite{M-FMA, FP-MPE}. Power consumption is lower than most evaluated units; only~\cite{Flex-PE} achieves lower power, at the cost of greater area and latency. No single competing design simultaneously achieves comparable area, power, and frequency, underscoring the multi-objective efficiency of the proposed DQ-MAC unit.

\subsection{Evaluation metrics for accelerator performance}

To comprehensively evaluate the proposed TREA accelerator, we define four hardware-efficiency metrics: placement complexity, cycle-level efficiency, inference latency, and energy consumption.

\begin{itemize}
    \item \textbf{Normalised Fabric Placement Complexity Index (nFPCI):} Reflects FPGA placement and routing complexity through the combined utilisation of LUT-6 and FDRE/FDCE cells, computed as:
    \[
    \text{nFPCI} = \left(\frac{\text{L}}{\text{L}_{\text{total}}}\right)\left(\frac{\text{FF}}{\text{FF}_{\text{total}}}\right)\times 1000,
    \]
    where L and FF denote the utilized LUT-6 and FDRE/FDCE resources, and L$_{\text{total}}$, FF$_{\text{total}}$ the corresponding total available resources on the target device. Lower values indicate a more resource-efficient design.

    \item \textbf{Clock Cycles per Inference Frame (CPFI):} The total number of clock cycles required to process a single input frame, providing a frequency-independent measure of cycle-level efficiency. Lower values indicate higher throughput efficiency.

    \item \textbf{Single-Frame Inference Latency (SFIL):} The total time required to process a single input frame, computed as:
    \[
    \text{SFIL} = \frac{\text{CPFI}}{f_{\text{clk}}}
    \]
    where $f_{\text{clk}}$ is the accelerator's operating clock frequency. Lower values indicate reduced end-to-end latency.

    \item \textbf{Energy Cost per Inference (ECPI):} The total energy (in $\mu$J) consumed per input frame. Lower values indicate improved energy efficiency and extended operational lifetime under fixed power budgets.

\end{itemize}

\subsection{Architectural Evaluation}

\begin{figure}[!t]
    \centering
    \includegraphics[width=\columnwidth]{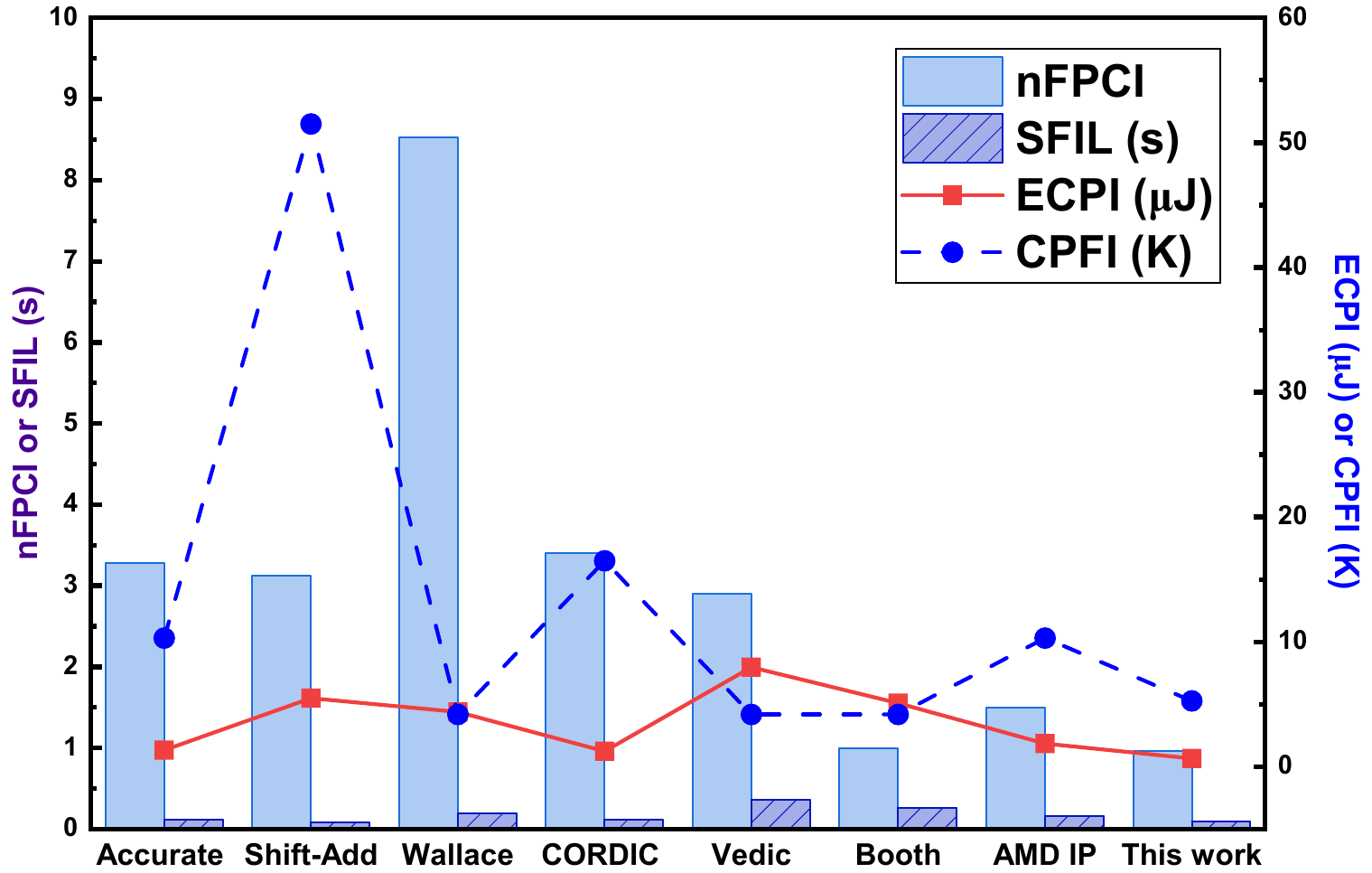}
    \caption{Performance comparison of FPGA metrics for TREA accelerator approach with conventional 8-bit MAC units.}
    \label{fig:trea_convmac}
\end{figure}

The suitability of the DQ-MAC unit as the processing element within the proposed accelerator is evaluated by substituting several state-of-the-art MAC designs into the TREA architecture and comparing system-level performance across key metrics, with results presented in Fig.~\ref{fig:trea_convmac}. The DQ-MAC achieves among the lowest nFPCI values, indicating strong resource efficiency. While the Booth multiplier-based accelerator yields a lower nFPCI, the DQ-MAC outperforms it across ECPI and SFIL, demonstrating superior energy efficiency and inference latency. The DQ-MAC achieves the lowest SFIL across all evaluated designs while maintaining favourable ECPI. The only comparable design in ECPI is CORDIC-based; however, its substantially higher nFPCI and SFIL values indicate greater resource complexity and lower throughput. Overall, the DQ-MAC achieves the best balance among resource utilisation, energy efficiency, and inference latency, confirming its suitability as the processing element of choice within the proposed accelerator.

\begin{figure}[!t]
    \centering
    \includegraphics[width=\columnwidth]{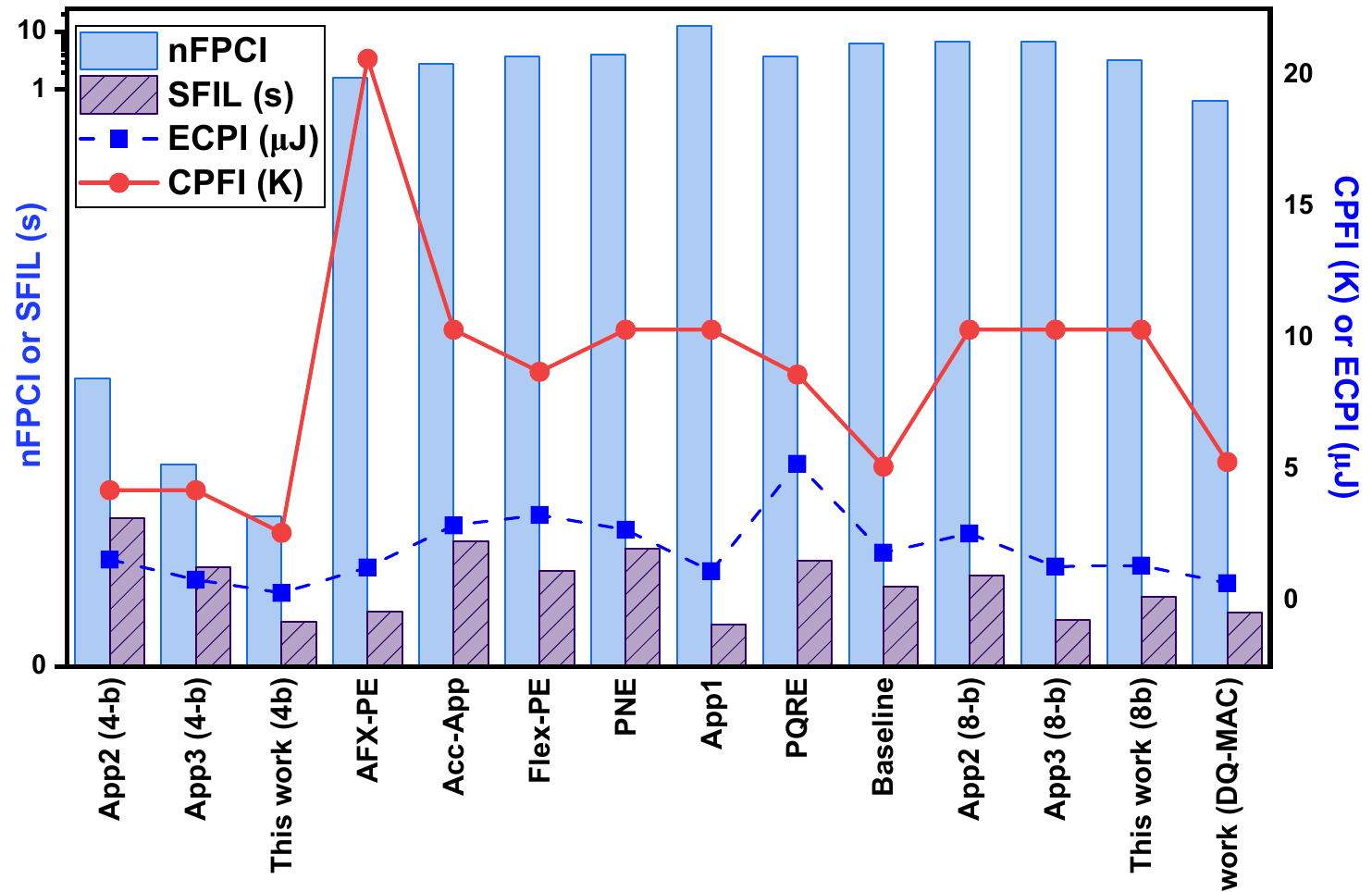}
    \caption{Performance comparison of FPGA metrics for TREA accelerator approach with prior 4-/8-bit MAC units\cite{LPRE, Acc-App-PE, Flex-PE, PNE, AppMAC_TETC'24, AppMAC}.}
    \label{fig:trea_sotamac}
\end{figure}

The comparison is extended to include both 4-bit and 8-bit state-of-the-art MAC designs, with accelerator-level results presented in Fig.~\ref{fig:trea_sotamac}. The first three configurations correspond to two representative 4-bit designs and the proposed 4-bit DQ-MAC; the DQ-MAC achieves favourable resource efficiency, energy efficiency, and inference latency across all evaluated metrics. Consistent trends are observed in the 8-bit comparison. The final configuration employs the DQ-MAC in reconfigurable precision mode, leveraging per-layer quantisation to achieve the lowest values across all metrics, yielding the highest throughput and best energy efficiency among all evaluated designs. Taken together, these results confirm the DQ-MAC as a well-suited compute element for the TREA accelerator.

\subsection{Comparison with SOTA work}

The proposed TREA accelerator is evaluated at the system level against state-of-the-art DNN accelerators across resource utilisation (LUTs, registers, DSPs), operating frequency, energy efficiency (GOPS/W), and power consumption, with results summarised in Table~\ref{tab:fpga-arch-results}. The proposed accelerator achieves one of the lowest resource footprints among evaluated designs, attributable to the DQ-MAC unit, the compute array organisation, and the adopted CORDIC implementation. This area efficiency is accompanied by a modest increase in power consumption relative to~\cite{RAMAN-IoTJ24} and~\cite{Wu-Sp-systolic-TCASI24}; however, the proposed design operates at a higher frequency, yielding comparable overall power-frequency performance.

\begin{table*}[!t]
\caption{FPGA Resource Utilisation for different domain-specific accelerator architectures, with resource enhancements in this work.}
\label{tab:fpga-arch-results}
\renewcommand{\arraystretch}{1.35}
\resizebox{\textwidth}{!}{%
\begin{tabular}{|l|c|c|c|c|c|c|c|c|c|c|c|}
\hline
\textbf{} & \textbf{\begin{tabular}[c]{@{}c@{}}TCAD'23\\ \cite{WJiang-TCAD'23} \end{tabular}} & \textbf{\begin{tabular}[c]{@{}c@{}}TVLSI'20\\ \cite{Zhu-SparseCNN-TVLSI20} \end{tabular}} & \textbf{\begin{tabular}[c]{@{}c@{}}TCAS-I'22\\ \cite{DThanh-TCASI'22} \end{tabular}} & \textbf{\begin{tabular}[c]{@{}c@{}}ESL'24\\ \cite{Yin-StrucSparse-ESL24} \end{tabular}} & \textbf{\begin{tabular}[c]{@{}c@{}}TVLSI'23\\ \cite{VaPr-WLee-TVLSI'23}\end{tabular}} & \textbf{\begin{tabular}[c]{@{}c@{}}TCAS-II'23\\ \cite{SKi-TCASII'23}\end{tabular}} & \textbf{\begin{tabular}[c]{@{}c@{}}IoTJ'24\\ \cite{RAMAN-IoTJ24} \end{tabular}} &
\textbf{\begin{tabular}[c]{@{}c@{}}TCAS-I'24\\ \cite{Wu-Sp-systolic-TCASI24}\end{tabular}} & \textbf{\begin{tabular}[c]{@{}c@{}}TVLSI'25\\ \cite{Flex-PE}\end{tabular}} &
\textbf{\begin{tabular}[c]{@{}c@{}}Access'24\\ \cite{QuantMAC} \end{tabular}} & \textbf{Proposed} \\ \hline

\textbf{FPGA Platform} & ZCU-102 & ZCU102 & KCU15 & ZCU102 & ZCU102 & XCVU9P & Ti60 & XC7Z035 & VC707 & VC707 & VC707 \\ \hline
\textbf{Workload} & MobileNet-v2 & VGG-16 & YoloV3-tiny & MobileNetV2 & XoR-Net & TinyYoloV3 & DS-CNN & VGG-16 & Mobile-ViT & VGG-16 & TinyYoloV3 \\ \hline
\textbf{Quant. Precision} & 8 & 16 & 8 & 16 & 1/2/4 & 8 & 2/4/8 & 8 & 4/8/16 & 8 & 4/8 \\ \hline
\begin{tabular}[c]{@{}l@{}}\textbf{LUT-6 slices (M)}\\ \color[HTML]{009901} Savings (x times) \end{tabular} & \begin{tabular}[c]{@{}c@{}}0.17\\ \color[HTML]{009901} (-5.67x)\end{tabular} & {\begin{tabular}[c]{@{}c@{}}\color[HTML]{FE0000} \textbf{0.41}\\ \color[HTML]{009901} (-13.67x)\end{tabular}} & {\begin{tabular}[c]{@{}c@{}}\color[HTML]{FE0000} \textbf{0.24}\\ \color[HTML]{009901} (-8.0x)\end{tabular}} & {\begin{tabular}[c]{@{}c@{}}\color[HTML]{FE0000} \textbf{0.22}\\ \color[HTML]{009901} (-7.34x)\end{tabular}} & \begin{tabular}[c]{@{}c@{}}0.18\\ \color[HTML]{009901} (-6.0x)\end{tabular} & \begin{tabular}[c]{@{}c@{}}0.13\\ \color[HTML]{009901} (-4.34x)\end{tabular}  & 0.04 & 0.04 & 0.04 & 0.04 & {\color[HTML]{32CB00} \textbf{0.03}} \\ \hline

\begin{tabular}[c]{@{}l@{}}\textbf{FDRE/FDCE (M)}\\ \color[HTML]{009901} Savings (x times)\end{tabular} & - & { \begin{tabular}[c]{@{}c@{}}\color[HTML]{FE0000} \textbf{0.28}\\ \color[HTML]{009901} (-14x)\end{tabular}} & {\begin{tabular}[c]{@{}c@{}}\color[HTML]{FE0000} \textbf{0.35}\\ \color[HTML]{009901} (-17.5x)\end{tabular}} & \begin{tabular}[c]{@{}c@{}}0.11\\ \color[HTML]{009901} (-5.5x)\end{tabular} & \begin{tabular}[c]{@{}c@{}}0.08\\ \color[HTML]{009901} (-4x)\end{tabular} & 0.04  & {\color[HTML]{32CB00} \textbf{0.01}} & 0.05 & \begin{tabular}[c]{@{}c@{}}0.14\\ \color[HTML]{009901} (-7x)\end{tabular} & 0.05 & {\color[HTML]{32CB00} \textbf{0.02}} \\ \hline

\textbf{DSP Blocks (K)} & {\color[HTML]{FE0000} \textbf{1.28}} & {\color[HTML]{FE0000} \textbf{1.35}} & {\color[HTML]{FE0000} \textbf{2.24}} & 0.89 & 0.14 & 0.13 & 0.12 & 0.26 & {\color[HTML]{32CB00} 0} & 0.04 & {\color[HTML]{32CB00} 0} \\ \hline

\begin{tabular}[c]{@{}l@{}}\textbf{Freq. (GHz)}\\ \color[HTML]{009901} Enhanced (x times)\end{tabular} 
& {\color[HTML]{32CB00} \textbf{0.33}} 
& 0.20 
& 0.20 
& \begin{tabular}[c]{@{}c@{}}0.19\\ \color[HTML]{009901} (-1.32x)\end{tabular} 
& 0.20 
& {\begin{tabular}[c]{@{}c@{}}\color[HTML]{FE0000} \textbf{0.10}\\ \color[HTML]{009901} (-2.5x)\end{tabular}} 
& {\begin{tabular}[c]{@{}c@{}}\color[HTML]{FE0000} \textbf{0.08} \\ \color[HTML]{009901} (-3.2x)\end{tabular}} 
& \begin{tabular}[c]{@{}c@{}}0.15\\ \color[HTML]{009901} (-1.67x)\end{tabular} 
& \begin{tabular}[c]{@{}c@{}}0.12\\ \color[HTML]{009901} (-2.1x)\end{tabular} 
& \begin{tabular}[c]{@{}c@{}}0.14\\ \color[HTML]{009901} (-1.8x)\end{tabular} 
& {\color[HTML]{32CB00} \textbf{0.25}} \\ \hline

\begin{tabular}[c]{@{}l@{}}\textbf{Peak PD (W)}\\ \color[HTML]{009901} Savings (x times)\end{tabular} & 0.96 & {\begin{tabular}[c]{@{}c@{}}\color[HTML]{FE0000} \textbf{15.50}\\ \color[HTML]{009901} (-9.7x)\end{tabular}} & {\color[HTML]{32CB00} \textbf{0.51}} & {\begin{tabular}[c]{@{}c@{}}\color[HTML]{FE0000} \textbf{13.30}\\ \color[HTML]{009901} (-8.3x)\end{tabular}} & \begin{tabular}[c]{@{}c@{}}06.58\\ \color[HTML]{009901} (-4.1x)\end{tabular} & 02.00 & {\color[HTML]{32CB00} \textbf{0.14}} & 01.40 & \begin{tabular}[c]{@{}c@{}}02.24\\ \color[HTML]{009901} (-1.4x)\end{tabular} & 01.80 & 01.60 \\ \hline

\begin{tabular}[c]{@{}l@{}}\textbf{Power Perf. (GOPS/W)}\\ \color[HTML]{009901} Enhanced (x times)\end{tabular} & \begin{tabular}[c]{@{}c@{}}05.20\\ \color[HTML]{009901} (+7.7x)\end{tabular} & \begin{tabular}[c]{@{}c@{}}07.80\\ \color[HTML]{009901} (+5.1x)\end{tabular} & \begin{tabular}[c]{@{}c@{}}10.30\\ \color[HTML]{009901}(+3.9x)\end{tabular} & \begin{tabular}[c]{@{}c@{}}23.50\\ \color[HTML]{009901}(+1.7x)\end{tabular} & \begin{tabular}[c]{@{}c@{}}12.90\\ \color[HTML]{009901}(+3.1x)\end{tabular} & \begin{tabular}[c]{@{}c@{}}04.50\\ \color[HTML]{009901}(+8.9x)\end{tabular} & {\color[HTML]{32CB00} \textbf{66.70}} & {\color[HTML]{32CB00} \textbf{45.00}} & \begin{tabular}[c]{@{}c@{}}08.42\\ \color[HTML]{009901}(+4.75x)\end{tabular} & {\begin{tabular}[c]{@{}c@{}} \color[HTML]{FE0000} \textbf{0.68}\\ \color[HTML]{009901} (+58.82x)\end{tabular}} & {\color[HTML]{32CB00} \textbf{40.00}} \\ \hline

\end{tabular}}
\end{table*}

\section{Conclusion}
This work introduced TREA, a time-multiplexed and resource-efficient edge-AI accelerator that leverages dual-precision SIMD computation, structured pruning, and reconfigurable activation hardware to address the performance–energy–area constraints of edge vision workloads. The proposed DQ-MAC unit achieves up to 4x throughput improvement through 4-bit SIMD execution while eliminating multiplier overhead via MSDF-based shift-and-add computation.

The SHARP pruning strategy enables nearly 50\% structured sparsity with full hardware utilisation, reducing kernel execution latency from 9 to 1 cycle for 3x3 filters and from 25 to 3 cycles for 5x5 filters, resulting in up to 9x latency improvement.
The integration of a 9-stage pipelined CORDIC-based activation unit provides one output per cycle with (N-1) hardware reuse, significantly reducing area overhead compared to conventional per-layer activation implementations.

By combining time-multiplexed layer reuse with a 100-unit SIMD MAC array, TREA minimises redundant hardware, reduces control complexity, and improves overall resource utilisation. The proposed architecture demonstrates substantial gains in latency, throughput, and energy efficiency compared to fixed-precision and non-reconfigurable designs, making it a scalable and practical solution for real-time object detection and classification on resource-constrained edge platforms.

\bibliographystyle{ieeetr}
\bibliography{thisbib}

\end{document}